\documentclass[a4paper,11pt]{article}
\usepackage{jheppub} 
\usepackage{lineno}

\usepackage{amsmath,amssymb}
\usepackage{amsthm}
\usepackage{amsfonts}
\usepackage{enumerate}
\usepackage{ytableau}
\usepackage{hyperref}
\usepackage{bm}
\usepackage{tikz}
\usepackage{blkarray}
\usepackage{multirow}
\usepackage{booktabs}
\usepackage{graphicx}
\usepackage{cite}
\usepackage{float}
\usepackage{simplewick}
\usepackage{arydshln}

\begin{document}

%===========================redefine================================
\def\bb#1\ee{\begin{equation}\begin{split}#1\end{split}\end{equation}}
\def\s#1{|_{#1}}
\def\({\left(}
\def\){\right)}
\def\[{\left[}
\def\]{\right]}
\def\be{\begin{equation}}
\def\ee{\end{equation}}

\def\L{L}
\def\Lb{\bar{L}}
\def\e{\operatorname{e}}
\def\vx{\vec{x}}
\def\TT{T\bar{T}}
\def\tr{\operatorname{tr}}
\def\pa{\partial}

\makeatletter 
\@addtoreset{equation}{section}
\makeatother
\renewcommand{\theequation}{\arabic{section}.\arabic{equation}}

\vspace{0.7cm} 
\begin{center}
\Large{\textbf{Self-Dual Electrodynamics via the Characteristic Method: Relativistic and Carrollian Perspectives}}
\vspace{1.3cm}

\normalsize{\textrm{Bin Chen$^{1,2}$, Song He$^{1,3}$, Jue Hou$^{4,5}$}}
\end{center}
\par~\par

\begin{center}
\footnotesize{\textit{
$^{1}$Institute of Fundamental Physics and Quantum Technology,\\
\& School of Physical Science and Technology,\\
Ningbo University, Ningbo, Zhejiang 315211, China\\
$^{2}$School of Physics, \& Center for High Energy Physics, Peking University, \\No.5 Yiheyuan Rd, Beijing 100871, P. R. China\\
$^{3}$Max Planck Institute for Gravitational Physics (Albert Einstein Institute),\\ Am M$\ddot{u}$hlenberg 1, 14476 Golm, Germany\\
$^{4}$School of Physics \& Shing-Tung Yau Center, Southeast University, Nanjing 211189, P. R. China\\
$^{5}$Mathematics Department, King's College London, The Strand, London WC2R 2LS, United Kingdom
}}
\par~\par

\footnotesize{\textit{Corresponding authors' emails:} chenbin1@nbu.edu.cn, hesong@nbu.edu.cn, juehou@seu.edu.cn}\\

\vspace{2cm}

\textbf{Abstract}\vspace{2mm}
\end{center}

\noindent
Electric-magnetic duality plays a pivotal role in understanding the structure of nonlinear electrodynamics (NED). The Gaillard-Zumino (GZ) criterion provides a powerful constraint for identifying self-dual theories. In this work, we systematically explore solutions to the GZ self-duality condition by applying the method of characteristics, a robust tool for solving nonlinear partial differential equations. Our approach enables the construction of new classes of Lagrangians that respect duality symmetry, both in the relativistic and Carrollian frameworks. In the relativistic setting, we not only recover well-known examples such as Born-Infeld and ModMax theories, but also identify novel models. We then generalize the GZ formalism to the Carrollian case and construct several classes of Carrollian self-dual non-linear electrodynamic models. Remarkably, we demonstrate that the characteristic flow exhibits an attractor behavior, in the sense that different seed theories that may not be self-dual can generate the same descendant self-dual Lagrangian. These findings broaden the landscape of self-dual theories and open new directions for exploring duality in ultra-relativistic regimes.

\vspace{2cm}

{\it This article is dedicated to Professor Ke Wu in Capital Normal University in celebration of his 80th birthday.}

\setcounter{page}{1}
\renewcommand{\thefootnote}{\arabic{footnote}}
\setcounter{footnote}{0}
\setcounter{tocdepth}{2}
\newpage
\tableofcontents

\section{Introduction}

It has been known for a long time that the four-dimensional Maxwell theory without sources exhibits invariance under discrete electric-magnetic duality, $\vec{E} \leftrightarrow \vec{B}$. Later, it was realized that the Maxwell theory is, in fact, invariant under continuous $SO(2)$ rotations\cite{rainich1925electrodynamics}:
\begin{equation}\label{SO2rotation} 
\vec{E} + i\vec{B} \to e^{-i\theta}(\vec{E} + i\vec{B}).
\end{equation}
The Maxwell theory is not unique in $SO(2)$ self-duality--several nonlinear electrodynamics (NED) also possess this property\cite{Gibbons:1995cv}. For $SO(2)$ self-dual nonlinear electrodynamics, the Lagrangian $\L(F)$ must satisfy the Gaillard-Zumino criterion~\cite{GAILLARD1981221,Gaillard:1997rt,Gibbons:1995ap,Hatsuda:1999ys}(see also \cite{Kuzenko:2000uh,Banerjee:2003bp, Aschieri:2008ns,Kallosh:2011dp,Bossard:2011ij,Carrasco:2011jv,Chemissany:2011yv,Aschieri:2013nda}):
\begin{equation}\label{GZ}
    G_{\mu\nu}\tilde{G}^{\mu\nu} + F_{\mu\nu}\tilde{F}^{\mu\nu} = 0,
\end{equation}
where $\tilde{F}, \tilde{G}$ can be defined in the following forms:
\bb\label{relationGF}
\tilde{G}_{\mu \nu}=\frac{\partial \L}{\partial S}F_{\mu\nu}+\frac{\partial \L}{\partial P}\Tilde{F}_{\mu\nu},\quad G_{\mu \nu}=-\frac{\partial \L}{\partial S}\Tilde{F}_{\mu\nu}+\frac{\partial \L}{\partial P}F_{\mu\nu}.
\ee
It is straightforward to show that  
\bb\label{GZCondition}
\((\partial_S \L)^2-(\partial_P \L)^2-1\)P-2 \partial_S \L\partial_P \L S=0.
\ee
where $S\equiv \frac{1}{4}F_{\mu\nu}F^{\mu\nu}$ and $ P\equiv\frac{1}{4}F_{\mu\nu}\Tilde{F}^{\mu\nu}$.
In addition to electric-magnetic $SO(2)$ duality, the Maxwell theory possesses conformal invariance, manifested through the tracelessness of its energy-momentum tensor:
\bb\label{traceT}
T^\mu{}_\mu=4\(S \partial_S \L+ P \partial_P \L -\L\)=0.
\ee
Equation~\eqref{GZCondition} admits various solutions for nonlinear electrodynamics (NED) Lagrangians. General methods for solving it were discussed in ~\cite{cour}, while perturbative solutions were addressed in ~\cite{Kuzenko:2000uh, Carrasco:2013qia}. Alternative and novel approaches can be found in ~\cite{Avetisyan:2021heg,Mkrtchyan:2022ulc,Ferko:2023wyi, Russo:2024llm, He:2025ppz,Babaei-Aghbolagh:2020kjg,Babaei-Aghbolagh:2022itg}.

In the Gaillard-Zumino framework, Maxwell's equations possess an electric-magnetic $SO(2)$ duality symmetry. However, extending this symmetry to the level of the action becomes nontrivial, especially in the presence of nonlinear interactions. Though the Lagrangians themselves are not $SO(2)$-invariant, their derivatives with respect to duality-invariant parameters, such as coupling constants or background fields like the spacetime metric, do exhibit the symmetry~\cite{GAILLARD1981221}.
The self-duality condition in this approach ensures that physical observables, including the equations of motion~\cite{Green:1996qg}, energy-momentum tensor~\cite{Gibbons:1995ap, BabaeiVelni:2016qea}, and scattering amplitudes~\cite{Babaei-Aghbolagh:2013hia, Garousi:2017fbe, Pavao:2022kog, BabaeiVelni:2019ptj}, remain invariant under S-duality transformations.

Recent developments have identified the \textit{ModMax theory} as the unique $SO(2)$ self-dual extension of the Maxwell theory that satisfies both the Gaillard-Zumino criterion and the tracelessness condition~\cite{Bandos:2020jsw, KOSYAKOV2020135840, Ferko:2022iru, Conti:2022egv, Ferko:2022cix, Hou:2022csf}\footnote{While the Bialynicki-Birula theory also meets these criteria, it lacks a Lagrangian formulation.}.
Relaxing the tracelessness condition admits other $SO(2)$ self-dual models, including the \textit{Born-Infeld electrodynamics}~\cite{Born:1933lls, Born:1934gh} and the \textit{generalized ModMax framework}~\cite{Bandos:2020hgy, Sorokin:2021tge,Babaei-Aghbolagh:2022MoxMax}. The constraints on the Lagrangian of such theories have been explored perturbatively in~\cite{Babaei-Aghbolagh:2024uqp}, while additional constructive methods for solving the Gaillard-Zumino criterion have appeared in~\cite{Mkrtchyan:2022ulc, Mkrtchyan:2019opf, Russo:2024llm, Russo:2024ptw, Ferko:2024yhc}.

Electric-magnetic duality can also be studied in electrodynamical theories that do not preserve Lorentz symmetry. A particularly compelling setting is \textit{Carrollian field theory}, which arises as the ultra-relativistic limit of Poincar\'e symmetry~\cite{Levy-Leblond:1965,SenGupta:1966qer}. Carrollian symmetry has recently attracted considerable interest for its fundamental role in diverse physical systems~\cite{Duval:2014uoa,Duval:2014uva,Duval:2014lpa,Hansen:2020hrs,Hao:2021urq,Chen:2021xkw,Yu:2022bcp,Hao:2022xhq,Banerjee:2022ocj,Bagchi:2022eui,Chen:2023pqf,Chen:2023esw,Chen:2023naw,Ecker:2023uwm,Mason:2023mti,Bergshoeff:2023vfd,Ciambelli:2023xqk,Kasikci:2023zdn,Marsot:2023qlc,deBoer:2023fnj,Nguyen:2023miw,Aggarwal:2024yxy,Stieberger:2024shv,Bagchi:2024unl,Liu:2024nkc,He:2024yzx,Tadros:2024qlo,Banerjee:2024jub}.

Carrollian electromagnetism exhibits distinct structural features, with two inequivalent limits: the \textit{electric-type} and \textit{magnetic-type} Carrollian Maxwell theories~\cite{Duval:2014uoa,Bagchi:2016bcd,Basu:2018dub,Banerjee:2020qjj,deBoer:2021jej,Henneaux:2021yzg,Bergshoeff:2022qkx,Banerjee:2022sza,Chen:2023pqf,deBoer:2023fnj}. Notably, the electric-type Carrollian Maxwell theory exhibits self-duality under a discrete EM duality transformation, although it lacks continuous $SO(2)$ duality invariance. The Carrollian Born-Infeld theory has been explored in~\cite{Mehra:2024zqv}, and a recent study in~\cite{Chen:2024vho} uncovers intriguing duality properties. By adapting the Gaillard-Zumino criterion to the Carrollian framework, a unique Carrollian ModMax theory has been constructed~\cite{Chen:2024vho}, which satisfies both:
\begin{itemize}
  \item the Carrollian version of the Gaillard-Zumino self-duality condition,
  \item the traceless energy-momentum tensor requirement.
\end{itemize}
This theory serves as a Carrollian counterpart to the relativistic ModMax model, preserving essential duality properties in the ultra-relativistic limit. Further exploration of the Carrollian Gaillard-Zumino equation may yield additional self-dual solutions.

The main motivation of this work is to find more self-dual theories in both relativistic and Carrollian electromagnetic frameworks. This requires us to search for general solutions to the Gaillard-Zumino (GZ) criterion, which governs the conditions under which nonlinear electrodynamic theories preserve electric-magnetic duality symmetry.  In this work, we undertake a systematic approach to this problem by employing the method of characteristics, a powerful tool in the analysis of nonlinear partial differential equations, to probe the space of nonlinear Lagrangians compatible with self-duality. Our method allows us to reinterpret the GZ criterion as a system of first-order PDEs, whose characteristic surfaces encode the propagation of field configurations respecting duality invariance. By solving these equations, we uncover a rich variety of self-dual theories, including several novel models in the Carrollian limit. This unified framework not only sheds light on the structure of duality-preserving theories but also paves the way for exploring their applications in holography, ultra-relativistic limits, and potentially quantum gravity.

The remaining parts of this paper are organized as follows. Section \ref{section2} reviews the Gaillard-Zumino (GZ) criterion for self-duality in nonlinear electrodynamics and shows how to solve it by using the method of characteristics.  In particular, we demonstrate how to generate self-dual theories from seed theories that are generically not self-dual. In Section \ref{section3}, we examine the solution of self-dual theories from the solutions of seed theories. In Section \ref{section4}, we extend the investigations to the Carrollian case. We summarize our results and future directions in Section \ref{section5}.

\section{Solving the Gaillard-Zumino equation via the method of characteristics}\label{section2}

To uncover novel self-dual theories in nonlinear electrodynamics, it is essential to systematically solve the Gaillard-Zumino (GZ) self-duality condition, which manifests as a nonlinear first-order partial differential equation for the Lagrangian. 
Mathematically, the GZ criterion manifests as a nonlinear, first-order partial differential equation (PDE) for the Lagrangian $L(S, P)$, where $S$ and $P$ are the two standard Lorentz invariants constructed from the field strength tensor $F_{\mu\nu}$. In this section, we present a unified, geometric approach for solving the GZ equation using the characteristics method. We show how this method enables exact and perturbative constructions of self-dual Lagrangians and introduce a classification scheme for the resulting families of duality-invariant actions.

\subsection{Method of characteristics}\label{sectionchar}
This section is briefly devoted to the method of characteristics. For the first-order fully nonlinear differential equation~\cite{evans2010partial,Hou:2022csf},
\begin{equation}
\left\{
\begin{array}{l}
F(\vec{x},u,\partial_{\vec{x}}{u})=0, \quad \vec{x} \in \mathbb{R}^{n} ,\\
u|_{\Gamma}=\L_0,
\end{array}
\right.
\end{equation}
where $\vec{x}$ is a collection of $n$ variables and $\Gamma$ is a $(n-1)$-dimensional manifold in $\mathbb{R}^{n}$. Let us parameterize $\Gamma$ by a $(n-1)$ dimension vector $\vec{r}=(r_1,...,r_{n-1})$, so that $\Gamma=(x_1(\vec{r},s),...,x_n(\vec{r},s))|_{s=0}\equiv(\gamma_1(\vec{r}),...,\gamma_n(\vec{r}))$. Defining $p_{i}(\vec{x}(\vec{r},s))=\partial_{x_{i}}u(\vec{x}(\vec{r},s))$, where $s$ is the affine parameter of the characteristic curve, the differential equation becomes
\begin{equation}\begin{split}\label{Nonliear first order PDE}
F(\vec{x},u,\vec{p})=0.
\end{split}\end{equation}
We want to find a vector field tangent to any integral surface defined by (\ref{Nonliear first order PDE}). The integral surfaces are unions of integral curves along the vector field, a.k.a. characteristic curves. For a vector field to be tangent to an integral surface, we must have
\bb
\frac{d F}{d s}=\frac{\partial F}{\partial \vec{x}}\cdot \frac{d\vec{x}}{ds} + \frac{\partial F}{\partial u}\frac{du}{ds} + \frac{\partial F}{\partial \vec{p}}\cdot \frac{d\vec{p}}{ds} &= 0 , 
\ee
and by the chain rule,
\begin{align}
    &\frac{du(\vec{x}(\vec{r},s))}{ds} = \frac{\partial u}{\partial \vec{x}}\cdot \frac{d\vec{x}}{ds}=\vec{p}\cdot \frac{d\vec{x}}{ds},\\
    &\frac{d\vec{p}(\vec{x}(\vec{r},s))}{ds}=\frac{\partial \vec{p}}{\partial \vec{x}}\cdot  \frac{d\vec{x}}{ds}.
\end{align}
The first term of the last equation depends on the particular integral surface. To get rid of this term, we differentiate (\ref{Nonliear first order PDE}) with respect to $\vec{x}$ for an integral surface
\begin{align}
    \frac{\partial F}{\partial \vec{x}} + \frac{\partial F}{\partial u} \vec{p} + \frac{\partial F}{\partial \vec{p}} \frac{\partial \vec{p}}{\partial \vec{x}} = 0.
\end{align}
We find the dependence on the particular integral surface can be eliminated if we set $\frac{d\vec{x}}{ds} = \frac{\partial F}{\partial\vec{p}}$. Then, we obtain the equation for characteristic curves of nonlinear first-order PDEs
\begin{align}
\frac{d\vec{x}}{ds} &= \frac{\partial F}{\partial \vec{p}} ,\\
    \frac{du}{ds} &= \vec{p}\cdot\frac{\partial F}{\partial \vec{p}} ,\\
\frac{d\vec{p}}{ds} &= -\frac{\partial F}{\partial\vec{x}} - \frac{\partial F}{\partial u}\vec{p}.
\end{align}
More apparently, we can introduce a set of $2n+1$ characteristic equations by
\begin{equation}\label{charaeq}
\begin{aligned}
\frac{d x_{i}(\vec{r}, s)}{d s} &=\frac{\partial F}{\partial {p_{i}}} ,\\
\frac{d u(\vec{r}, s)}{d s} &=\sum_{i=1}^{n} p_{i} \frac{\partial F}{\partial {p_{i}}}  ,\\
\frac{d p_{i}(\vec{r}, s)}{d s} &=-\frac{\partial F}{\partial {x_{i}}} -\frac{\partial F}{\partial u}  p_{i},
\end{aligned}
\end{equation}
with the boundary conditions 
\begin{equation}
\begin{aligned}
&x_{i}(\vec{r}, 0)=\gamma_{i}(\vec{r}) ,\quad u(\vec{r}, 0)=u_0(\vec{r}) ,\quad p_{i}(\vec{r}, 0)=\psi_{i}(\vec{r}), \quad \vec{r} \in \mathbb{R}^{n-1},
\end{aligned}
\end{equation}
where $i=1,2,...,n$. The $n$ unknown functions $\psi_i(\vec{r})$ satisfy
\begin{equation}\label{eqboundary}
\begin{aligned}
&\frac{\partial u_0}{\partial r_i}=\psi_1(\vec{r}) \frac{\partial \gamma_{1}}{\partial r_{i}}+\ldots+\psi_{n}(\vec{r}) \frac{\partial \gamma_{n}}{\partial r_{i}}, \quad i=1, \ldots, n-1 ,\\
&F\left(\gamma_{1}(\vec{r}), \ldots, \gamma_{n}(\vec{r}), u_0(\vec{r}), \vec{\psi}_1(\vec{r}), \ldots, \psi_{n}(\vec{r})\right)=0.
\end{aligned}
\end{equation}
It is worth emphasizing that the solution of $\psi_i(\vec{r})$ may not exist or be unique. If we get the solution of characteristic equations, $(\vec{x}(\vec{r}, s), u(\vec{r}, s), \vec{p}(\vec{r}, s))$, and can find the inverse functions of $\vec{x}(\vec{r}, s), \vec{p}(\vec{r}, s)$, then we can eliminate $\vec{r},s$ in $u(\vec{r}, s)$ to get $u(\vec{x})$.

\subsection{Rewriting the GZ criterion}

We take the variables
\begin{equation}
S, \quad P, \quad L, \quad p_S = \frac{\partial L}{\partial S}, \quad p_P = \frac{\partial L}{\partial P}.
\end{equation}
Equation~\eqref{GZCondition} can then be expressed more compactly as
\begin{equation}
F(S, P, L, p_S, p_P) = (p_S^2 - p_P^2 - 1)P - 2p_S p_P S = 0. \label{eq:GZreduced}
\end{equation}
This relation defines a constraint hypersurface in the extended space of variables $(S, P, p_S, p_P)$. Our goal is to determine all functions $L(S, P)$ that satisfy this constraint.

The method of characteristics transforms the first-order PDE~\eqref{eq:GZreduced} into a system of ordinary differential equations (ODEs) that describe the evolution of $(S, P, L, p_S, p_P)$ along a parameter $s$, referred to as the flow parameter. The characteristic equations associated with~\eqref{eq:GZreduced} are
\begin{subequations} \label{charaeq2}
\begin{align}
\label{charaeqa}\frac{d S}{d s} &=2 p_S P-2 p_P S,\\
\label{charaeqb}\frac{d P}{d s} &=-2 p_P P-2 p_S S,\\
\label{charaeqc}\frac{d L}{d s} &=2 P,\\
\label{charaeqd}\frac{d p_S}{d s} &=2 p_S p_P,\\
\label{charaeqe}\frac{d p_P}{d s} &=-\(p_S^2-p_P^2-1\).
\end{align}
\end{subequations}

These equations are subject to initial (or boundary) conditions specified along a seed curve:
\begin{align}
S(0) &= \gamma_S(r), \quad P(0) = \gamma_P(r), \quad L(0) = L_0(r), \nonumber \\
p_S(0) &= \psi_S(r), \quad p_P(0) = \psi_P(r),
\end{align}
where $r$ is a parameter along the initial data curve. The functions $\psi_S(r)$ and $\psi_P(r)$ must satisfy the constraint:
\begin{equation}
\frac{dL_0}{dr} = \psi_S \frac{d\gamma_S}{dr} + \psi_P \frac{d\gamma_P}{dr}, \quad \text{and} \quad F|_{s=0} = 0.
\end{equation}

Finally, the solutions of the characteristic equations~\eqref{charaeq2} can be expressed as follows 
\bb\label{solx1x2z}
S=&\frac{1}{2}  \left(\gamma_S \left(\psi_S^2-\psi_P^2+1\right)+2 \gamma_P \psi_S \psi_P\right)\(\cos (2 s)-1\)+(\gamma_P \psi_S-\gamma_S \psi_P)\sin (2 s)+\gamma_S,\\
P=&\gamma_P \cos(2s)-\(\psi_S \gamma_S+\psi_P\gamma_P\) \sin(2s),\\
L=& \L_0+\(\psi_S \gamma_S+\psi_P\gamma_P\)\(\cos(2s)-1\)+ \gamma_P\sin(2s).
\ee
Here $\gamma_i, \psi_i, \L_0$ satisfy \eqref{eqboundary}
\bb\label{eqboundary2}
\frac{\partial \L_0}{\partial r}&=\psi_S\frac{\partial \gamma_{1}}{\partial r}+\psi_{2}\frac{\partial \gamma_{2}}{\partial r},\\
F|_{s=0}&=\(\psi_S^2-\psi_P^2-1\)\gamma_P-2 \psi_S \psi_P \gamma_S=0,
\ee
where $r,s$ are the parameters of the characteristic surface and the boundary surface (curve) $\Gamma$ is $(r,s=0)$. To obtain Lagrangians of $SO(2)$ self-dual theories, we need to set a seed theory  $\L_0$ and the boundary values $\gamma_i, \psi_i$ satisfying \eqref{eqboundary2}, and eliminate the parameter $s$ in $L$ by $S, P$. It is worth noting that the seed theory can't be a isolated fixed point along the characteristic flow. We will explain the process by examples in the following subsections.

\subsection{Examples and recovering known theories} 

To illustrate the practical utility of the characteristic method in solving the Gaillard-Zumino self-duality condition, we present a set of concrete examples, which are the solutions to the characteristic flow equations. These examples include known self-dual Lagrangians, such as Maxwell and ModMax theories, and more intricate nonlinear electrodynamics models, including the Born-Infeld theory and its Carrollian counterparts.
These examples validate our method and highlight its potential to uncover novel duality-invariant models in both relativistic and ultra-relativistic regimes. 

\paragraph{Type I: Polynomial Seeds.}
To obtain exact analytical expressions for the resulting actions, we impose boundary conditions on $\gamma_i$, $\psi_i$, and $\L_0$ for $i=1,2$. We aim to classify all analytically tractable self-dual Lagrangians, which can be expressed in closed form. For clarity, we list only representative non-vanishing boundary conditions that reproduce known models in the literature.
\begin{equation}
\psi_S(r) = \sum_{i=0}^{n}{a_i r^i}.
\end{equation}
Here $a_i$ is the coefficients of the series expansion.
Since in our situation, we can only obtain the analytical solution for $n\leq 1$, we will consider the two simplest seed Lagrangians, $\L_0 = - a r$ and $\L_0 = - a r + b r^2$, where $a$ and $b$ are constants along the characteristic flow. These constants may depend on external parameters but remain independent of the characteristic variables $S, P, s$.

\subsubsection*{Example 1: ModMax}

We first consider the seed Lagrangian:
\bb
\gamma_S = r, \quad \gamma_P = 0, \quad \L_0 = - a r, \quad \text{with } a\neq 1.
\ee
Applying equation \eqref{eqboundary2}, we obtain:
\bb
\psi_S = -a, \quad \psi_P = 0.
\ee
Substituting these boundary values into \eqref{solx1x2z}, we derive:
\bb
&S = \frac{1}{2} r \left(1 - a^2 + (1 + a^2) \cos(2s) \right),\quad P = r a \sin(2s),
\quad L= -r a \cos(2s).
\ee
Eliminating $r$ and $s$ in favor of $S$ and $P$, we find:
\bb
L= \frac{-(a^2+1)S \pm (a^2-1) \sqrt{S^2 + P^2}}{2a}.
\ee
This corresponds to the Lagrangian of the ModMax. Rewriting it in the standard form by setting $a = e^{\alpha}$, we obtain the resulting Lagrangian:
\bb
L &= -\cosh(\alpha) S \pm \sinh(\alpha) \sqrt{S^2 + P^2}\nonumber
\ee
Thus, the resulting action coincides with the standard ModMax action \cite{Bandos:2020jsw}.

It is worth emphasizing that for the seed Lagrangian, we require $a\neq 1$. If $a=1$, the seed theory and the descendant theory become same and are both Maxwell. The Maxwell is self-dual and so it is a fixed point along the characteristic flow.

\subsubsection*{Example 2} 

For this case, the seed Lagrangian is given by:
\bb
\gamma_S = r, \quad \gamma_P = 0, \quad \L_0 = - a r + b r^2.
\ee
Applying equation \eqref{eqboundary2}, we obtain:
\bb
\psi_S = -a + 2 b r, \quad \psi_P = 0.
\ee
Substituting these boundary values into \eqref{solx1x2z}, we derive:
\bb
S &= \frac{1}{2} r \left(1 - (a - 2 b r)^2 + \left(1 + (a - 2 b r)^2\right) \cos(2s) \right),\\
P &= r (a - 2 b r) \sin(2s),\\
L &= -r \left((a - 2 b r) + b r\right) \cos(2s).
\ee
Eliminating $r$ and $s$ in favor of $S$ and $P$ yields the Lagrangian $L$. However, the resulting relations form a sixth-order equation in $r$, preventing an explicit solution in terms of elementary functions. To proceed, we set $a = e^{\alpha}$ and treat $b$ as a small parameter, expanding the Lagrangian perturbatively in powers of $b$:
\bb
L=& -\cosh(\alpha)S + \sinh(\alpha)\sqrt{S^2 + P^2}\\
&+ \frac{1}{2} e^{-2\alpha} \left( P^2 + (2S^2 + P^2) \cosh(2\alpha) - 2S \sqrt{S^2 + P^2} \sinh(2\alpha) \right) b + O(b^2).
\ee
Since the sixth-order equation has multiple solutions, the above Lagrangian represents only a perturbative approximation.

This model lacks conformal invariance, which can be verified by computing the trace of the energy-momentum tensor, $T^{\mu}{}_{\mu}$. A more direct approach is to use equations \eqref{charaeqc}, \eqref{charaeqd} and \eqref{charaeqe}, which imply that $p_S S + p_P P - L$ remains constant along the characteristic flow. From equation \eqref{traceT}, we obtain the trace relation:
\bb
T^{\mu}{}_{\mu} = 4(p_S S + p_P P - L) = 4(\psi_S \gamma_S + \psi_P \gamma_P - \L_0).
\ee
Thus, if the energy-momentum tensor of the seed theory is not traceless, neither is the one along the characteristic flow, and vice versa. In this case, it is straightforward to check that the energy-momentum tensor of the seed theory is not traceless.

\par~\par
\subsection{New Classes of Self-Dual Theories}\label{sec2} 

We seek new $SO(2)$ self-dual theories with a closed-form Lagrangian. For simplicity, we adopt the same boundary values as before:
\bb
\gamma_S = r, \quad \gamma_P = 0.
\ee
From equation \eqref{eqboundary2}, we obtain:
\bb
\psi_P = 0, \quad \psi_S = \frac{\partial \L_0}{\partial r}.
\ee  
Thus, we refer to the seed theory interchangeably as $\psi_0$ or $\L_0$. Simplifying equation \eqref{solx1x2z}, we derive:
\bb
S &= \frac{1}{2} r \left( 1 - \psi_S^2 + (1 + \psi_S^2) \cos(2s) \right),\\
P &= -r \psi_S \sin(2s),\\
L &= \L_0 + r \psi_S \left( \cos(2s) - 1 \right).
\ee
Eliminating $s$, we obtain:
\bb\label{expressionz}
L= \L_0 + \frac{2 \psi_S}{1 + \psi_S^2} (S - r),
\ee
where $r$ satisfies:
\bb\label{eqaboutr}
P^2 + (-4 r S + 4 S^2 + 2 P^2) \psi_S^2 + (-4 r^2 + 4 r S + P^2) \psi_S^4 = 0.
\ee
As seen in the previous subsection, for $\psi_S = -a + 2 b r$, $r$ satisfies a sixth-order polynomial equation, precluding a closed-form elementary solution. To obtain a Lagrangian expressible in elementary functions, careful selection of the seed theory is required. We identify several suitable seed theories, leading to new families of exact NED models.

\subsubsection{Type II} 

We choose the seed theory as:
\bb
\psi_S = \sqrt{\frac{a}{b + c r + d r^2}},
\ee
where $a, b, c, d$ are constants. The resulting theory is denoted as Type \uppercase\expandafter{\romannumeral1}-$(a; b, c, d)$. For $d \neq 0$, integrating $r$ in $\psi_S$ yields:
\bb\label{type1Bc}
\L_0 = \frac{\sqrt{a} \log \left(2 \left(\sqrt{d (b + r (c + d r))} + d r\right) + c\right)}{\sqrt{d}} + \text{const}.
\ee
Since only three of $a, b, c, d$ are independent, we can set $d = 1$ by rescaling $a, b, c$. Substituting the boundary conditions into \eqref{expressionz} and \eqref{eqaboutr}, we obtain $SO(2)$ self-dual Lagrangians. If $d = 0$, the analysis simplifies significantly. Below, we examine specific cases.

\subsubsection*{Type \texorpdfstring{\uppercase\expandafter{\romannumeral2}}{2} -\texorpdfstring{$(a;b,c,0)$}{(a;b,c,0)}} 
There are four solutions,
\begin{subequations}\label{case1sol}
\begin{align}
\label{case1sol1}&\L=\pm\sqrt{\left(2 a+S+\sqrt{S^2+P^2}\right) \left(2 b+S-\sqrt{S^2+P^2}\right)}+const,\\
\label{case1sol2}&\L=\pm\sqrt{\left(2 a+S-\sqrt{S^2+P^2}\right) \left(2 b+S+\sqrt{S^2+P^2}\right)}+const.
\end{align}\end{subequations}

\begin{itemize}
    \item One of the solutions can become the other three by analytic continuation. That is to say, the four solutions are actually different representations of one function in different sheets of the Riemann surface.

    \item These solutions are self-dual in the following sense. Taking the $S>0, P>0$ branch for example, 
    \bb
    L=-\sqrt{\left(2 a+S-\sqrt{S^2+P^2}\right) \left(2 b+S+\sqrt{S^2+P^2}\right)}+const.
    \ee 
    Under the transformation,
    \bb
    S \rightarrow \sqrt{-S^2 - P^2},\quad P \rightarrow P,\quad a \rightarrow - i a,\quad b\rightarrow i b,
    \ee
the Lagrangian is invariant, so it is a discrete symmetry of the Lagrangian. However, it is worth noting that the Lagrangian will change under the $SO(2)$ self-dual transformation \eqref{SO2rotation}. The $SO(2)$ self-dual transformation only preserves the equation of motion but not the Lagrangian.
\end{itemize}

Let us illustrate this type of solution with a few specific examples. We select one of four solutions and restore the dependence on the parameter $c$,
\bb\label{case1solr}
L=-\sqrt{\left( \frac{2a}{c}+S-\sqrt{S^2+P^2}\right) \left( \frac{2b}{c}+S+\sqrt{S^2+P^2}\right)}+const,
\ee
\begin{itemize}
    \item Taking  $a=b=\frac{c}{4 \mu},const=\frac{1}{2\mu}$, with  $\mu>0$, the above Lagrangian becomes
    \bb
    L=\frac{1}{2\mu}\(1-\sqrt{1+4 \mu S-4\mu^2 P^2}\),
    \ee
    which is one of the Born-Infeld (BI) theories.
    \item Taking  $a=-\frac{c}{4 \mu e^\alpha},b=-\frac{c e^\alpha}{4\mu},const=\frac{1}{2\mu}$ with $\mu>0$, the above Lagrangian becomes
    \bb
    L=\frac{1}{2\mu}\(1-\sqrt{1+4 \mu \(-\cosh(\alpha)S+\sinh(\alpha)\sqrt{S^2+P^2}\)-4\mu^2 P^2}\)
    \ee
    This is the Generalized Born-Infeld (GBI) theory. 
    \item Taking $const=\frac{2 \sqrt{a b}}{c}, a=\frac{1}{4 e^\alpha}, b=\frac{e^\alpha}{4}$ and expanding the above Lagrangian around $c=0$, the $O(1/c)$ term cancels out. Under the $c \rightarrow 0$ limit, the Lagrangian becomes
    \bb
    L=-\cosh(\alpha)S+\sinh(\alpha)\sqrt{S^2+P^2},
    \ee
    which gives the ModMax theory. 
\end{itemize}
As it stands, the Born-Infeld, the generalized Born-Infeld theory, and the ModMax theory all belong to the same type! They have the same type of seed theories, $\psi_S=\sqrt{\frac{a}{b+c r}}$.

There is another interesting phenomenon. When $a=0$, the seed theory is trivial, $\psi_S=0, \L_0=const$, but the solution \eqref{case1solr} is not trivial 
\bb\label{sola0bc}
L=-\sqrt{\left( S-\sqrt{S^2+P^2}\right) \left( \frac{2b}{c}+S+\sqrt{S^2+P^2}\right)}+const.
\ee
This is a new theory satisfying the Gaillard-Zumino criterion.

If we take $a=b=0$, the solutions \eqref{case1solr} become $L=\pm i |P|$, that is, $\L=\pm  \frac{i}{4}|F_{\mu\nu}\Tilde{F}^{\mu\nu}|$, which is the strong field limit of the BI theory. If we transform the result to Hamiltonian form and then take $a=b=0$, we can arrive at the Bialynicki-Birula (BB) theory, whose Hamiltonian is given by $H_{BB}=|D\times B|$. The BB theory is also in the same type with the BI, the GBI and the ModMax.

In summary,  the BI, the GBI, the ModMax, the BB and the theories given by \eqref{sola0bc} all belong to Type \uppercase\expandafter{\romannumeral1}-$(a;b,c,0)$, whose Lagrangian is given by \eqref{case1solr}. Especially, the BI and the GBI belong to Type \uppercase\expandafter{\romannumeral1}-$(a;b,1,0)$, the ModMax belongs to Type \uppercase\expandafter{\romannumeral1}-$(a;b,0,0)$, the BB is Type \uppercase\expandafter{\romannumeral1}-$(0;0,0,0)$, and the theories given by \eqref{sola0bc} are of Type \uppercase\expandafter{\romannumeral1}-$(0;b,1,0)$\footnote{It also belongs to Type \uppercase\expandafter{\romannumeral1}-$(a;0,1,0)$, since Type \uppercase\expandafter{\romannumeral1}-$(0;b,1,0)$ and Type \uppercase\expandafter{\romannumeral1}-$(a;0,1,0)$ are related by analytic continuation.}.

\subsubsection*{Type \texorpdfstring{\uppercase\expandafter{\romannumeral2}}{2} -\texorpdfstring{$(a^2;b^2,2 b,1)$}{(a^2;b^2,2b,1)}}

We now consider a related class with boundary data:
\bb
\psi_S = \sqrt{\frac{a^2}{b^2 + 2 b r + r^2}} = \frac{a}{b + r}, \quad \L_0 = a \log(b + r) + \text{const}.
\ee
Substituting these boundary conditions into equations \eqref{expressionz} and \eqref{eqaboutr} yields multiple solutions, which are potentially related by analytic continuation.

One such solution takes the form:
\bb
L= \sqrt{a^2 + (S - \Omega)(2b + S + \Omega)} + a \log\left(\frac{a \left(\sqrt{a^2 + (S - \Omega)(2b + S + \Omega)} - a\right)}{S - \Omega} \right) - a + \text{const},
\ee
where $\Omega = \sqrt{S^2 + P^2}$.

A similar Lagrangian structure involving a logarithmic term has appeared in black hole solutions coupled to nonlinear electrodynamics and gravity, as discussed in \cite{Soleng:1995kn}.

\subsubsection*{Type \texorpdfstring{\uppercase\expandafter{\romannumeral2}}{2} -\texorpdfstring{$(a;b,0,1)$}{(a;b,0,1)}}

We now examine a class of theories defined by the boundary conditions:
\bb
\psi_S = \sqrt{\frac{a}{b + r^2}}, \quad \L_0 = \sqrt{a} \tanh^{-1}\left(\frac{r}{\sqrt{b + r^2}}\right) + \text{const}.
\ee
Substituting into equations \eqref{expressionz} and \eqref{eqaboutr}, we obtain the resulting Lagrangian:
\bb
L= -\sqrt{2a - P^2 - 2 \omega} + \sqrt{a} \tanh^{-1}\left(\frac{a - \omega}{\sqrt{a(2a - P^2 - 2 \omega)}}\right) + \text{const},
\ee
where $\omega = \sqrt{a^2 + a(S - \Omega)(S + \Omega) - b(S - \Omega)^2}$ and $\Omega = \sqrt{S^2 + P^2}$.

This type adds to the catalog of exact Lagrangian constructions derived from self-dual seed theories, enriching the structure of nonlinear electrodynamics with new functional forms.

\subsubsection{Type III}

We now turn to another class defined by the seed function:
\bb
\psi_S = \sqrt{\frac{a + b r}{c + d r}}.
\ee
The resulting Lagrangian $L$ can be expressed in terms of elementary functions. We denote this class as Type \uppercase\expandafter{\romannumeral2}-$(a,b;c,d)$.

As a specific example, consider Type \uppercase\expandafter{\romannumeral2}-$(a,b;1,0)$ with boundary data:
\bb
\psi_S = \sqrt{a + b r}, \quad \L_0 = \frac{2(a + b r)^{3/2}}{3b} + \text{const}.
\ee
Substituting into \eqref{expressionz} and \eqref{eqaboutr}, one solution takes the form:
\bb
L= \frac{\sqrt{2a + b(S + \Omega) + \omega}(4a + 2b(S + \Omega) - \omega)}{6b} + \text{const},
\ee
where $\omega = \sqrt{(2a + b S + b \Omega)^2 - 8b(S - \Omega)}$ and $\Omega = \sqrt{S^2 + P^2}$.

In the case of (anti-)self-duality, i.e., $F_{\mu\nu} = \pm \tilde{F}_{\mu\nu}$ so that $S^2 = P^2$, and setting $P^2 = S^2$ with $a = 2\sqrt{2} - 3$, the Lagrangian simplifies to:
\bb
L= \frac{\sqrt{\sqrt{2} + 1} \sqrt{S} \left[(\sqrt{2} + 2) b S - 18\sqrt{2} + 24\right]}{6\sqrt{b}}.
\ee

\subsubsection{Type IV}

We consider the seed theory:
\bb
\psi_S = \left(\frac{a}{b + c r + d r^2}\right)^{1/4}.
\ee
Integrating over $r$, we obtain:
\bb
\L_0 = \frac{\left(\frac{a d}{4 b d - c^2}\right)^{1/4} (c + 2 d r) \, _2F_1\left(\frac{1}{4}, \frac{1}{2}; \frac{3}{2}; \frac{(c + 2 d r)^2}{c^2 - 4 b d}\right)}{\sqrt{2} d} + \text{const}.
\ee
For this class of seed theories, $r$ can always be expressed in terms of elementary functions of $S$ and $P$, although the Lagrangian itself is not always elementary. However, for certain special cases, it reduces to an elementary function. For instance, considering Type \uppercase\expandafter{\romannumeral3} -$(a; b,1,0)$:
\bb
\psi_S = \left(\frac{a}{b + r}\right)^{1/4}, \quad \L_0 = \frac{4}{3} (b + r) \left(\frac{a}{b + r}\right)^{1/4}.
\ee
We obtain the resulting Lagrangian:
\bb
L= \frac{\sqrt{\omega+\Omega-S}}{6\sqrt{a}(4a+\omega+\Omega-S)}\bigg(4a(4b+4S+\omega)+(S-\Omega)(\omega+\Omega-S)\bigg) + \text{const},
\ee
where $\omega = \sqrt{(S-\Omega)^2+8a(2b+S +\Omega)}$ and $\Omega = \sqrt{S^2 + P^2}$.

\subsubsection{Type V}

For this category, we define the non-vanishing initial boundary conditions as:
\bb
\psi_S = \sqrt{\frac{a}{b + \sqrt{c + d r} + e r}}.
\ee
Integrating over $r$, we can obtain the seed theory as follows:
\bb
\L_0 = \frac{2 \sqrt{a}}{e} \sqrt{b + \sqrt{c + d r} + e r} - \frac{\sqrt{a} \sqrt{d} }{e^{3/2}}\tanh^{-1}\left(\frac{2 e \sqrt{c + d r} + d}{2 \sqrt{d} \sqrt{e} \sqrt{b + \sqrt{c + d r} + e r}}\right) + \text{const}.
\ee

As a particular case, we examine the boundary condition:
\bb
\psi_S = \sqrt{\frac{a}{b + \sqrt{c + r}}}.
\ee
Integrating over $r$, we obtain:
\bb
\L_0 = \frac{4}{3} \sqrt{a} \left(\sqrt{c + r} - 2 b\right) \sqrt{b + \sqrt{c + r}} + \text{const}.
\ee
Substituting this boundary condition into \eqref{expressionz} and \eqref{eqaboutr}, we obtain the corresponding solution:
\bb
L= \frac{(-8 a b + 2 S + \omega - 2 \Omega) }{6 a} \sqrt{4 a b - S + \omega + \Omega}+ \text{const},
\ee
where:
\bb
\omega = \sqrt{8 a^2 (2 c + S + \Omega) + 8 a b (\Omega - S) + (S - \Omega)^2}, \quad \Omega = \sqrt{S^2 + P^2}.
\ee
This provides a novel theory with a particularly simple Lagrangian.

Interestingly, while the seed theory does not satisfy the Gaillard--Zumino criterion and is not $SO(2)$ self-dual, the resulting theory exhibits full $SO(2)$ invariance. This highlights a geometric attractor mechanism: the flow governed by the characteristic equations converges to an $SO(2)$-invariant structure, independent of the initial boundary choice. Such behavior suggests a universality in the emergent symmetry of the final theory, even when it is absent in the seed formulation.

According to \eqref{eqaboutr}, a sixth-order polynomial equation, if we want the descendant theory with a closed form, we have to choose the seed theory carefully. For exFor example, we require that $\psi_S$ is an algebraic function with respect to $r$.is section, we list some seed theories whose descendant theories have a closed-form Lagrangian. We believe we have listed most of the closed-form Lagrangians, but there may still be other Lagrangians with good forms. 

Furthermore, by other methods, such as the Courant-Hilbert method \cite{Russo:2024llm, Babaei-Aghbolagh:2025uoz, Murcia:2025psi} and the auxiliary fields approach \cite{Russo:2025fuc}, one can also classify the self-dual theories. These methods deal with the same equation —the Gaillard-Zumino equation —so they are equivalent. These different methods can be regarded as different representations of a method. We show the equivalence in Appendix \ref{appA}.

\section{Characteristics flows vs \texorpdfstring{$SO(2)$}{SO(2)} self-duality}\label{section3}

The Gaillard-Zumino equation \eqref{GZCondition}
can be  rewritten  as
\bb
\epsilon^{\mu\nu\rho\sigma}F_{\mu\nu}F_{\rho\sigma}-\epsilon_{\mu\nu\rho\sigma}\Tilde{G}^{\mu\nu}\Tilde{G}^{\rho\sigma}=0,
\ee
Then the characteristic equations are given by
\bb\label{characteristicsFij}
&\frac{d F_{\mu\nu}}{d s}=-\frac{1}{2}\epsilon_{\mu\nu\rho\sigma}\Tilde{G}^{\rho\sigma}=G_{\mu\nu},\\
&\frac{d \Tilde{G}^{\mu\nu}}{d s}=-\frac{1}{2}\epsilon^{\mu\nu\rho\sigma}F_{\rho\sigma}=-F_{\mu\nu},\\
&\frac{d L}{d s}=\frac{1}{2}G_{\mu\nu}\Tilde{G}^{\mu\nu}.
\ee
The solution is
\bb\label{solFij}
&F_{\mu\nu}=\cos(s) f_{\mu\nu}+\sin(s) g_{\mu\nu},\\
&G_{\mu\nu}=-\sin(s) f_{\mu\nu}+\cos(s) g_{\mu\nu},\\
\ee
where $f_{\mu\nu}=F_{\mu\nu}(s=0), g_{\mu\nu}=G_{\mu\nu}(s=0)$. The solution turns out to be  $SO(2)$ self-dual transformation of the original electric-magnetic theory. Plugging the solution into the third equation of \eqref{characteristicsFij}, we find 
\bb\label{solFijz}
\L \equiv L=\L_0-\frac{1}{4} f_{\mu\nu}\Tilde{f}^{\mu\nu} \sin(2 s) +\frac{1}{4} f_{\mu\nu}\Tilde{g}^{\mu\nu}\(\cos(2 s)-1\),
\ee
where $\L_0=\L(s=0)$. We can rewrite the solution about the Lagrangian in a more symmetric form 
\bb
\L-\frac{1}{4} F_{\mu\nu}\Tilde{G}^{\mu\nu}=\L_0-\frac{1}{4} f_{\mu\nu}\Tilde{g}^{\mu\nu},
\ee
which indicates that  $\L-\frac{1}{4} F_{\mu\nu}\Tilde{G}^{\mu\nu}$ is invariant under the $SO(2)$ self-dual transformation \cite{Gaillard:1997rt}.

In Section \ref{section2}, we showed how to generate $SO(2)$ self-dual theories from general seed theories that could be not  $SO(2)$ self-dual. The descendant theory generally has a complicated Lagrangian and equations of motion (EOM) that are hard to solve. On the other hand, the seed theory is simple, and its EOM is easy to solve.  So it would be great if we could find a way to generate the solution of EOM of the descendant theory from the solution of EOM of the seed theory. As we showed above,  the characteristic flow is the $SO(2)$ rotation, see  \eqref{solFij}.
Therefore,  if we have a solution of EOM of the seed theory, such as $\bar{A}_{\mu}$, then plugging the solution into the relation \eqref{solFij}, we can get the solution of EOM of the descendant theory. In this section, we show how the process works with the ModMax.

Considering the ModMax, whose Lagrangian is given by
\bb\label{LagrangianMM}
\L=\frac{-(a^2+1)S+(a^2-1)\sqrt{S^2+P^2}}{2a}.
\ee
The EOM of the ModMax is complicated. However, we may get the ModMax from a simple model,
\bb
\L_0=-a S_0=\frac{a}{4}f_{\mu\nu}f^{\mu\nu},
\ee
which is almost the Maxwell with a simple EOM,
\bb\label{eqfmunu}
\partial_\mu f^{\mu\nu}=0,\\
\ee
and a constraint ensuring the existence of $A_\mu$,
\bb\label{eqftmunu}
\partial_\mu \Tilde{f}^{\mu\nu}=0.
\ee
The initial data includes
\bb
S(s=0)=\gamma_S=r,\quad P(s=0)=\gamma_P=0,\quad \L_0=- a r.
\ee
{The initial data on $P$ suggests that 
\bb
P(s=0)=-\frac{1}{4}f_{\mu\nu}\Tilde{f}^{\mu\nu}=0
\ee
which impose constraints on the initial field strength. We may 
 take $f_{01}, f_{02}, f_{03}, f_{12}, f_{13}$ and the characteristic parameter $s$ as the free initial variables and then 
\bb\label{constraintf}
f_{23}=\frac{f_{02}f_{13}-f_{03}f_{12}}{f_{01}}.
\ee
In the following, we show how to obtain the solutions of the EOM of the ModMax from the above input.}

Consider the boundary with $s=0$ and
\bb
f_{01},\quad f_{02},\quad f_{03},\quad f_{12},\quad f_{13},\quad f_{23}=\frac{f_{02}f_{13}-f_{03}f_{12}}{f_{01}}.
\ee
The Lagrangian of the seed theory is given by
\bb
\L_0=-\frac{a}{4}f_{\mu\nu}f^{\mu\nu}=\frac{a}{2}\(f^2_{01}+f^2_{02}+f^2_{03}-f^2_{12}-f^2_{13}-\frac{\(f_{02}f_{13}-f_{03}f_{12}\)^2}{f^2_{01}}\).
\ee
By \eqref{eqboundary2}, we get $\Tilde{g}^{\mu\nu}=\Tilde{G}^{\mu\nu}(s=0)$
\bb
&\Tilde{g}^{01}=a f_{01},\quad \Tilde{g}^{02}=a f_{02},\quad \Tilde{g}^{03}=a f_{03},\quad \Tilde{g}^{12}=-a f_{12},\\& \Tilde{g}^{13}=-a f_{13},\quad
\Tilde{g}^{23}=\frac{\Tilde{g}^{02}\Tilde{g}^{13}-\Tilde{g}^{03}\Tilde{g}^{12}}{\Tilde{g}^{01}}.
\ee
Plugging the boundary data into \eqref{solFij}, we get
\bb\label{solFf}
&F_{01}=\cos(s) f_{01}-a \sin(s) \frac{f_{02}f_{13}-f_{03}f_{12}}{f_{01}},\\
&F_{02}=\cos(s) f_{02}+a \sin(s) f_{13},\\
&F_{03}=\cos(s) f_{03}-a \sin(s) f_{12},\\
&F_{12}=\cos(s) f_{12}+a \sin(s) f_{03},\\
&F_{13}=\cos(s) f_{13}-a \sin(s) f_{02},\\
&F_{23}=\cos(s)\frac{f_{02}f_{13}-f_{03}f_{12}}{f_{01}}+a \sin(s)f_{01}.\\
\ee
Using and above relations and eliminating $f_{\mu\nu}$, $\Tilde{g}^{\mu\nu}$ and $s$ by $F_{\mu\nu}$ in \eqref{solFijz}, we get
\bb
\L=\frac{-(a^2+1)S+(a^2-1)\sqrt{S^2+P^2}}{2a},
\ee
where $S=\frac{1}{4}F_{\mu\nu}F^{\mu\nu}, P=\frac{1}{4}F_{\mu\nu}\Tilde{F}^{\mu\nu}$. It is consistent with \eqref{LagrangianMM}.

{Let us start from the following solutions to  \eqref{eqfmunu}, \eqref{eqftmunu} and \eqref{constraintf},}
\bb
&f_{01}(t,x,y,z)=c_1 y+c_2 x y +c_3,\quad f_{02}(t,x,y,z)=\frac{1}{2}(2 x c_1+x^2 c_2-y^2 c_2+c_4),\\
&f_{03}(t,x,y,z)=f_{12}(t,x,y,z)=f_{13}(t,x,y,z)=0.
\ee
Plugging the above initial solutions into \eqref{solFf}, we get
\bb\label{solFf2}
&F_{01}= (c_1 y+c_2 x y +c_3) \cos(s),\quad F_{02}=\frac{1}{2}(2 x c_1+x^2 c_2-y^2 c_2+c_4)\cos(s) ,\\
&F_{13}=-\frac{1}{2}(2 x c_1+x^2 c_2-y^2 c_2+c_4) a \sin(s),\quad F_{23}=(c_1 y+c_2 x y +c_3) a \sin(s),\\
&F_{03}=0,\quad F_{12}=0,\\
\ee
where $a$ is the constant in Lagrangian of the ModMax and $c_1,c_2,c_3,c_4, s$ are free constants. It can be checked that the solutions \eqref{solFf2} satisfy EOM of the ModMax $\partial_{\mu}\Tilde{G}^{\mu\nu}=0$ and the Bianchi identity $\partial_{\mu}\Tilde{F}^{\mu\nu}=0$ for all $c_1,c_2,c_3,c_4, s$.

Let us consider another example, the plane wave solution. For the $L$-direction plane wave, its field strength is given by 
\bb
f_{\mu\nu}=e^{i(k z-\omega t)}
 \left(
\begin{aligned}
    0&&E_0&& 0&&0\\
    -E_0&&0&& 0&&B_0\\
    0&&0&& 0&&0\\
    0&&-B_0&& 0&&0\\
\end{aligned}
 \right),
\ee
where $E_0=B_0$ is a constant and $k=\omega$ is a constant as well. Then the solution of the EOM of the ModMax is given by
\bb
F_{\mu\nu}=E_0 e^{i(k z-\omega t)}
 \left(
\begin{aligned}
    0&& \cos(s)&& a  \sin(s)&&0\\
    - \cos(s)&&0&& 0&& \cos(s)\\
    -  a  \sin(s)&&0&& 0&&a  \sin(s)\\
    0&&- \cos(s)&& -a  \sin(s)&&0\\
\end{aligned}
 \right),
\ee
The amplitudes of electric and magnetic fields are
\bb
|\vec{E}|=|\vec{B}|=\sqrt{|E_x|^2+|E_y|^2+|E_z|^2}=E_0 \sqrt{\cos^2(s)+a^2 \sin^2(s)}.
\ee
{That means that the amplitudes vary with} the ModMax parameter $a$. When $a=1$, the ModMax degenerates to the Maxwell, and the solution satisfies the EOM of the Maxwell.

The third example is the static pure electronic field, whose field strength is given by
\bb
f_{\mu\nu}=
 \left(
\begin{aligned}
    0&&E_0&& 0&&0\\
    -E_0&&0&& 0&&0\\
    0&&0&& 0&&0\\
    0&&0&& 0&&0\\
\end{aligned}
 \right).
\ee
Then the solution of the EOM of the ModMax is given by
\bb
F_{\mu\nu}=E_0 
 \left(
\begin{aligned}
    0&& \cos(s)&& 0&&0\\
    - \cos(s)&&0&& 0&& 0\\
   0&&0&& 0&&a  \sin(s)\\
    0&&0&& -a  \sin(s)&&0\\
\end{aligned}
 \right),
\ee
It is known that the $SO(2)$ duality can mix electric fields and magnetic fields. However, for $a=0$, the Lagrangian \eqref{LagrangianMM} degenerates to
\bb
\L=-S-\sqrt{S^2+P^2}
\ee
up to a rescaling factor, and the pure electric fields remain unchanged, but can't generate magnetic fields. The $SO(2)$ duality only changes the amplitudes of electric fields. On the contrary, when $a=\infty$, the Lagrangian becomes 
$
\L=-S+\sqrt{S^2+P^2}
$
and only a pure magnetic field appears.

In the last section, we treat $S$ and $P$ as variables that evolve along the characteristic flow. In this section, we take $F_{\mu\nu}$ as variables. Plugging the solution \eqref{solFij} and \eqref{solFijz} into the definition of $S=\frac{1}{4}F_{\mu\nu}F^{\mu\nu}$ and $P=\frac{1}{4}F_{\mu\nu}\Tilde{F}^{\mu\nu}$, one can get the solution \eqref{solx1x2z} directly. Therefore, the two kinds of variables are equivalent for Lagrangians. However, there are 6 independent components of $F_{\mu\nu}$, but only two independent variables for $S$ and $P$. So we can only obtain $S$ and $P$ from $F_{\mu\nu}$, but can't obtain $F_{\mu\nu}$ from $S$ and $P$. When considering the solution to the equation of motion, we should use this section method with $F_{\mu\nu}$. If we want the Lagrangian, using $S$ and $P$ is easier since there are fewer variables.

\section{Carrollian \texorpdfstring{$SO(2)$}{SO(2)} self-duality}
\label{section4}

In this section, we explore the structure of $SO(2)$ self-duality in the context of Carrollian electrodynamics, which arises as the ultra-relativistic limit of relativistic field theories. Carrollian theories have attracted growing interest due to their appearance in flat space holography, tensionless strings, and non-Lorentzian geometry. The electromagnetic sector, in particular, admits two distinct limiting behaviors-electric-type and magnetic-type Carrollian theories, each exhibiting unique symmetry properties. We investigate how the Gaillard-Zumino self-duality condition can be adapted to the Carrollian setting and examine its implications for constructing Carrollian analogues of self-dual nonlinear electrodynamics. In doing so, we identify new models, including a Carrollian extension of the ModMax theory, which preserve essential features such as duality invariance and traceless energy-momentum tensors.

In \cite{Chen:2024vho}, a family of Carrollian ModMax theories, which are invariant under
Carrollian $SO(2)$ electric-magnetic (EM) duality and conformal transformations have been constructed. The duality transformations are subtler, as one must carefully define the Hodge duality in Carrollian geometry. The similar Gaillard-Zumino consistency condition for EM duality of Carrollian nonlinear electrodynamics was derived in \cite{Chen:2024vho},
\bb\label{cGZCondition}
\(-(\partial_P \L)^2-1\)P-2 \partial_S \L\partial_P \L S=0,
\ee
where the Carrollian Hodge dual was used to define $P$. 
Defining the two Carroll invariants $S, P$ and the Lagrangian $\L$, the Gaillard-Zumino criterion is written as
\bb\label{CGZconditionF}
F=\(-p_P^2-1\)P-2 p_S p_P S=0,
\ee
where $p_S=\frac{\partial \L}{\partial S}$ and $p_P=\frac{\partial \L}{\partial P}$. Then, the characteristic equations become
\bb \label{Ccharaeq2}
\frac{d S}{d s} &=-2 p_P S,\quad \frac{d P}{d s} =-2 p_P P-2 p_S S,\quad \frac{d L}{d s} =2 P,\\
\frac{d p_S}{d s} &=2 p_S p_P,\quad \frac{d p_P}{d s} =p_P^2+1.
\ee
Here, we have used \eqref{CGZconditionF} in the above equations. Similar to  the relativistic case, we get the solutions of \eqref{Ccharaeq2} as
\bb\label{Csolx1x2z}
S=&\gamma_S\(1-\frac{1}{2}\(\psi_P^2-1\)\(\cos(2s)-1\)-\psi_P\sin(2s)\),\\
P=&\gamma_P \cos(2s)-\(\psi_S \gamma_S+\psi_P\gamma_P\) \sin(2s),\\
L=& \L_0+\(\psi_S \gamma_S+\psi_P\gamma_P\)\(\cos(2s)-1\)+ \gamma_P\sin(2s).
\ee
Here, $\gamma_i, \psi_i, \L_0$ satisfy \eqref{eqboundary}
\bb\label{Ceqboundary2}
&\frac{\partial \L_0}{\partial r}=\psi_S\frac{\partial \gamma_{1}}{\partial r}+\psi_{2}\frac{\partial \gamma_{2}}{\partial r},\\
&F(s=0)=\(-\psi_P^2-1\)\gamma_P-2 \psi_S \psi_P \gamma_S=0.
\ee
It is noticed that the last two solutions in \eqref{Csolx1x2z} are the same as the ones in the relativistic version, but the first solution in \eqref{Csolx1x2z} about $S$ is different.

Consider a special boundary condition, where $\gamma_S=r, \gamma_P=0$. Then by \eqref{Ceqboundary2}, we can get $\psi_S=\frac{\partial \L_0}{\partial r}, \psi_P=0$. Under the boundary condition, the solutions \eqref{Csolx1x2z} become
\bb
S=r\(1+\frac{1}{2}\(\cos(2s)-1\)\),\quad P=-\psi_S r \sin(2s),\quad L= \L_0+\psi_S r\(\cos(2s)-1\).
\ee
Eliminate $r$ in the above equations, we have
\bb\label{Cexpressionz}
&\(\frac{2 S}{r}-1\)^2+\(\frac{P}{\psi_S r}\)^2=1,\\
&L= \L_0+2 \psi_S \(S-r\).
\ee

We now explore several representative examples illustrating the Carrollian self-dual framework.

\subsubsection*{Example 1: Carrollian ModMax}

Consider the seed theory:
\bb
\L_0 = -a r, \quad \psi_S = -a.
\ee
Substituting into \eqref{Cexpressionz}, we derive:
\bb
L= -a S + \frac{P^2}{4 a S}.
\ee
This is the Carrollian analog of the ModMax theory, previously introduced in \cite{Chen:2024vho}. It retains a form of $SO(2)$ self-duality in the Carrollian limit. Remarkably, this model exhibits a well-defined Carrollian limit of the relativistic ModMax theory and connects to the Carrollian Maxwell theory in the strong-field regime:
\bb
\lim_{a \to \infty} \frac{L}{a} = -S.
\ee
This limiting behavior reproduces the Carrollian Maxwell Lagrangian, $L= -S$, illustrating a clear hierarchy between the Carrollian Maxwell and the Carrollian ModMax. This hierarchy is reminiscent of how the Born-Infeld theory encompasses the Maxwell in its weak-coupling regime.

\subsubsection*{Example 2: Square-root model}

We now consider a nonlinear seed theory:
\bb
\L_0 = 2b \sqrt{r + a}, \quad \psi_S = \frac{b}{\sqrt{r + a}}.
\ee
Using \eqref{Cexpressionz}, we find
\bb
L= \sqrt{(a + S)\left(4 b^2 - \frac{P^2}{S}\right)}.
\ee
This model yields a closed-form Lagrangian with manifest nonlinearity and Carrollian invariance. The square-root structure introduces an upper bound on $P$ for fixed $S$, analogous to causality constraints in relativistic theories. The requirement for real-valued Lagrangians imposes $P^2 < 4 b^2 S / (a + S)$, suggesting a natural field-strength bound in the Carrollian setting.

\subsubsection*{Example 3: Logarithmic model}

We next analyze a logarithmic seed theory
\bb
\L_0 = b \log(r + a), \quad \psi_S = \frac{b}{r + a}.
\ee
Substituting into \eqref{Cexpressionz}, we obtain
\bb
L= \sqrt{b^2 - \frac{(a + S)P^2}{S}} - b + b \log\left(\frac{2b}{P^2} \left(b S - \sqrt{b^2 S^2 - (a + S)S P^2}\right)\right).
\ee
This model exhibits logarithmic behavior at weak field strengths and nonlinear corrections at higher orders. It interpolates between the Carrollian Maxwell-like behavior and strongly deformed self-dual dynamics. The nested square root and logarithmic structure reflect intricate geometric constraints on the field configuration, indicating rich dynamics that merit further exploration.

These examples demonstrate the versatility of the Carrollian construction in generating a broad class of exactly solvable, nonlinear self-dual electrodynamic models. Each model admits a closed-form Lagrangian in terms of $S$ and $P$, encoding both the seed theory's structure and the underlying geometric flow. The Carrollian framework parallels and enriches the relativistic self-dual structure, offering new insights into non-Lorentzian field theories with duality invariance. Additional examples and generalizations can be incorporated as needed to further illustrate the breadth of this approach.

\section{Conclusion and Perspectives}\label{section5}

In this work, we constructed broad families of nonlinear electrodynamics theories satisfying the Gaillard-Zumino criterion for $SO(2)$ self-duality without imposing conformal invariance. Employing the method of characteristics, we developed a systematic approach to generate exact or perturbative Lagrangians from arbitrary seed theories. This approach reveals a rich geometric structure in the solution space of duality-invariant NEDs. We explicitly identified multiple new classes of exact self-dual theories, including generalizations of Born-Infeld, ModMax, and Bialynicki-Birula models. The Lagrangian of these classes encompasses a broad functional forms, from elementary to transcendental functions.

We then extended the method to the Carrollian case, and we derived Carrollian analogs of self-dual theories, including the Carrollian ModMax. The corresponding characteristic flows exhibit modified duality conditions consistent with the Carrollian Gaillard-Zumino constraint. Despite the lack of Lorentz symmetry, the Carrollian models inherit duality-invariance through appropriately defined Hodge duals, highlighting a robust structural continuity between relativistic and Carrollian regimes.

The method of characteristics provides a technical means of solving the Gaillard-Zumino equation and reveals a deeper geometric structure underlying self-dual electrodynamics. In this framework, the GZ condition is treated as a constraint surface in a phase space defined by the field strength invariants $(S, P)$ and the derivatives of the Lagrangian $(\partial_S L, \partial_P L)$. The characteristic equations describe the evolution of field configurations along curves---so-called characteristic flows---that lie on this constraint surface. These flows are governed by a vector field derived from the GZ equation and trace out surfaces in the extended configuration space. The final form of the Lagrangian is obtained by eliminating the flow parameter and expressing the solution purely in terms of the original field invariants. This process is geometric: the self-dual Lagrangian corresponds to the scalar potential defined on the surface traced by the characteristics, constrained to obey duality invariance.

\begin{figure}[t]
\centering
\includegraphics[width=0.7\textwidth]{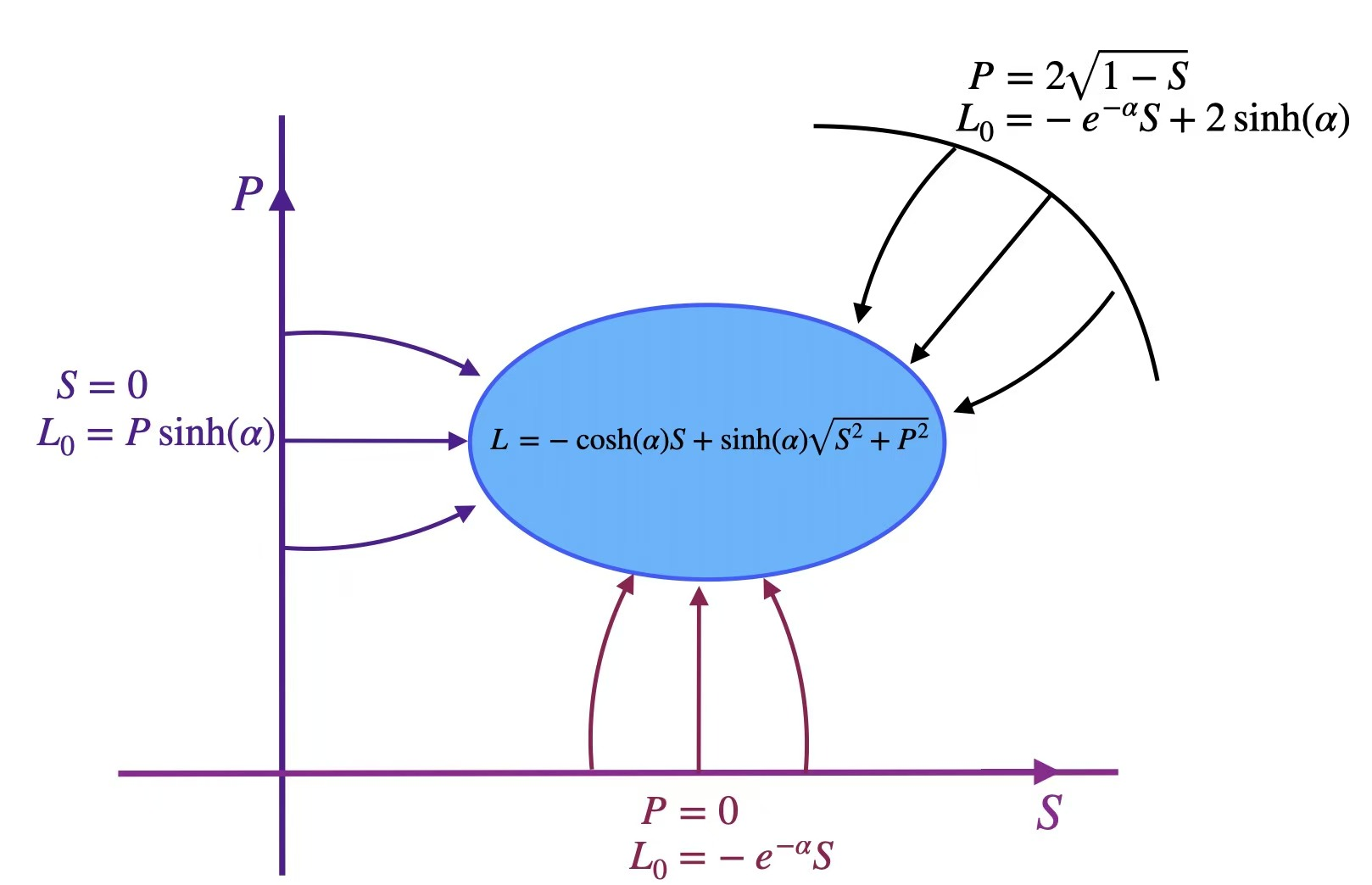}
\caption{Schematic of characteristic flow converging to a duality-invariant surface in $(S, P)$ space. That different boundary choices can generate the same self-dual Lagrangian. For instance, all of the boundary $P=0, L_0=-e^{-\alpha}S$, the boundary $S=0, L_0= \sinh(\alpha)P$, and the boundary $P=2 \sqrt{1-S}, L_0=  -e^{-\alpha}S +2 \sinh(\alpha)$ generate the ModMax.}
\label{fig:flow}
\end{figure}

One striking insight is that different boundary choices for the characteristic method---i.e., different ``seed'' theories or initial surfaces---can generate the same self-dual Lagrangian. For example, choosing boundary conditions aligned along the $S$-axis or the $P$-axis can yield the same solution up to an analytic continuation (see Figure \ref{fig:flow}). This implies that the characteristic flow exhibits an attractor-like behavior, where the long-term evolution becomes insensitive to the precise form of the seed theory, provided it satisfies the boundary constraints of the GZ equation. In this sense, self-duality emerges as an attractor on the $S-P$ plane (see Figure \ref{fig:flow}), suggesting that duality-invariant theories may be geometrically preferred or stable configurations in the broader functional space of electrodynamics Lagrangians. This attractor mechanism hints at universality in the structure of nonlinear electrodynamics, where different theories converge to the same duality-invariant outcome despite distinct origins. It opens a window for exploring deeper connections between symmetry, geometry, and dynamics in field theory.

Our results suggest several directions for future exploration. First, the emergence of duality-invariant structures from non-self-dual seeds suggests a possible attractor mechanism in the functional space of NEDs. Clarifying this geometric behavior could shed light on broader questions of universality and symmetry restoration in effective field theories. Second, while we focused on abelian gauge fields, the formalism may be adaptable to non-abelian settings or higher $p$-form theories. Lastly, applying this framework to holography, black hole thermodynamics\footnote{For example, in \cite{Barbosa:2025smt}, authors explore all kinds of non-linear electrodynamic models about the weak gravity conjecture.}, or double-trace deformations in string-inspired settings remains a promising area for further investigations.

\section*{Acknowledgments}
We are grateful to H. Babaei-Aghbolagh, Christian Ferko, Tommaso Morone, Shanming Ruan, Gabriele Tartaglino-Mazzucchelli, and Roberto Tateo for helpful discussions and comments. We also thank the anonymous reviewer for the valuable comments about the Courant-Hilbert approach and the auxiliary fields approaches. This research was supported in part by NSFC Grant No.11735001, 12275004, 12475053, 12235016, 12588101.

\appendix
\section{Equivalence among different approaches}
\label{appA}
In the appendix, we derive the equivalence among the method of characteristics, the Courant-Hilbert method, and the auxiliary fields approaches for the Gaillard-Zumino equation.

\subsection{From characteristics to Courant-Hilbert}

The Gaillard-Zumino equation is given by \eqref{GZCondition}
\bb
\((\partial_S \L)^2-(\partial_P \L)^2-1\)P-2 \partial_S \L\partial_P \L S=0.
\ee
Introducing new variables,
\bb
x=\frac{i}{2}\(S+\sqrt{S^2+P^2}\),\quad y=\frac{i}{2}\(S-\sqrt{S^2+P^2}\),
\ee
the Gaillard-Zumino equation becomes
\bb\label{floweqF}
0=F=\frac{\pa z}{\pa x}\frac{\pa z}{\pa y}+1 \equiv p_x p_y+1,
\ee
where $p_x=\frac{\pa z}{\pa x}, p_y=\frac{\pa z}{\pa y}$. 
By the method of characteristics, the characteristic equations are given by
\begin{subequations} \label{charaeqs}\begin{align}
\label{charaeqsa}&\frac{d x}{d s}=\frac{\pa F}{\pa p_x}=p_y,\\
\label{charaeqsb}&\frac{d y}{d s}=\frac{\pa F}{\pa p_y}=p_x,\\
\label{charaeqsc}&\frac{d z}{d s}=p_x \frac{\pa F}{\pa p_x}+p_y \frac{\pa F}{\pa p_y}=2 p_x p_y=-2,\\
\label{charaeqsd}&\frac{d p_x}{d s}=\frac{d p_y}{d s}=0.
\end{align}
\end{subequations}
The solution is 
\bb\label{solxy}
&p_x=const=\psi_x,\quad p_y=const=\psi_y,\\
&x=r_x+s \psi_y,\quad y=r_y+s \psi_x,\quad z=z_0-2s.
\ee
The initial condition satisfies
\bb
\psi_x \psi_y+1=0.
\ee
Taking the boundary at $r_x=r, r_y=0$, then we have $\psi_x=\frac{\pa z_0}{\pa r}\equiv \dot{z}_0,\psi_y=-\frac{1}{\psi_x}$. Plugging them into \eqref{solxy}, we get
\bb
x=r-s\frac{1}{\dot{z}_0},\quad y=s \dot{z}_0,\quad z=z_0-2s.
\ee
Eliminating $s$, finally we obtain that
\bb
z=z_0-2\frac{y}{\dot{z}_0}, \quad r=x+\frac{y}{\dot{z}_0^2}.
\ee
We use the new notation $z=\mathcal{L}, z_0=\ell, x=q_2, y=q_1, r=\tau$, then the above equation becomes
\begin{equation}\label{CHframework}
\mathcal{L}=\ell(\tau)-\frac{2 q_1}{\dot{\ell}(\tau)}, \quad \tau=q_2+\frac{q_1}{\dot{\ell}^2(\tau)},
\end{equation}
which is the Courant-Hilbert equation shown in \cite{Russo:2024llm,Babaei-Aghbolagh:2025uoz,Murcia:2025psi}.

\par~\par

\subsection{From Courant-Hilbert to the auxiliary fields}

The equivalence between the Courant-Hilbert method and the auxiliary field frameworks has been shown in \cite{Babaei-Aghbolagh:2025uoz}. The new auxiliary-field formulation is fully equivalent to the Courant-Hilbert (CH) representation in terms of the generating function $\ell(\tau)$, with the two related by a simple Legendre transform. Denoting the auxiliary variable by $y=e^{\varphi}$ and its potential by $\Omega(y)$ (so that $W(\varphi)\equiv \Omega(y)$), the map reads
\begin{equation}
\ell(\tau)=\tau\,y-\Omega(y),\qquad 
\tau=\Omega'(y),\qquad 
y=\ell'(\tau),
\label{eq:legendre}
\end{equation}
which establishes a bijection between solutions of the CH equation written in terms of $\ell(\tau)$ and the stationary solutions of the auxiliary Lagrangian written in terms of $\Omega(y)$. In particular, the universal deformation identities take the same form on both sides,
\begin{eqnarray}
\partial_\gamma \mathcal{L}=\mathcal{R}_\gamma=\tau\,\ell'(\tau)=y\,\Omega'(y), 
\qquad
\partial_\lambda \mathcal{L}=\mathcal{O}_\lambda&=&-\ell(\tau)\!\left(\ell(\tau)-2\tau\ell'(\tau)\right)\nonumber\\
&=& -\Omega(y)^2+y^2\Omega'(y)^2,
\end{eqnarray}
making the equivalence operational for flow analyses.

\par~\par
\bibliographystyle{unsrt}
\bibliography{ref}

@article{GAILLARD1981221,
title = {Duality rotations for interacting fields},
journal = {Nuclear Physics B},
volume = {193},
number = {1},
pages = {221-244},
year = {1981},
issn = {0550-3213},
doi = {https://doi.org/10.1016/0550-3213(81)90527-7},
url = {https://www.sciencedirect.com/science/article/pii/0550321381905277},
author = {Mary K. Gaillard and Bruno Zumino},
}

@book{evans2010partial,
  title={Partial Differential Equations},
  author={Evans, Lawrence C},
  year={2010},
  publisher={American Mathematical Society}
}

@article{Kuzenko:2000uh,
    author = "Kuzenko, Sergei M. and Theisen, Stefan",
    title = "{Nonlinear selfduality and supersymmetry}",
    eprint = "hep-th/0007231",
    archivePrefix = "arXiv",
    reportNumber = "LMU-TPW-00-19",
    doi = "10.1002/1521-3978(200102)49:1/3<273::AID-PROP273>3.0.CO;2-0",
    journal = "Fortsch. Phys.",
    volume = "49",
    pages = "273--309",
    year = "2001"
}

@article{Kallosh:2011dp,
    author = "Kallosh, Renata",
    title = "{$E_{7(7)}$ Symmetry and Finiteness of N=8 Supergravity}",
    eprint = "1103.4115",
    archivePrefix = "arXiv",
    primaryClass = "hep-th",
    reportNumber = "SU-ITP-2011-09",
    doi = "10.1007/JHEP03(2012)083",
    journal = "JHEP",
    volume = "03",
    pages = "083",
    year = "2012"
}

@article{Bossard:2011ij,
    author = "Bossard, Guillaume and Nicolai, Hermann",
    title = "{Counterterms vs. Dualities}",
    eprint = "1105.1273",
    archivePrefix = "arXiv",
    primaryClass = "hep-th",
    reportNumber = "AEI-2011-026, CPHT-RR039.0511",
    doi = "10.1007/JHEP08(2011)074",
    journal = "JHEP",
    volume = "08",
    pages = "074",
    year = "2011"
}

@article{Carrasco:2011jv,
    author = "Carrasco, John Joseph M. and Kallosh, Renata and Roiban, Radu",
    title = "{Covariant procedures for perturbative non-linear deformation of duality-invariant theories}",
    eprint = "1108.4390",
    archivePrefix = "arXiv",
    primaryClass = "hep-th",
    reportNumber = "SU-ITP-11-41",
    doi = "10.1103/PhysRevD.85.025007",
    journal = "Phys. Rev. D",
    volume = "85",
    pages = "025007",
    year = "2012"
}

@article{Chemissany:2011yv,
    author = "Chemissany, Wissam and Kallosh, Renata and Ortin, Tomas",
    title = "{Born-Infeld with Higher Derivatives}",
    eprint = "1112.0332",
    archivePrefix = "arXiv",
    primaryClass = "hep-th",
    doi = "10.1103/PhysRevD.85.046002",
    journal = "Phys. Rev. D",
    volume = "85",
    pages = "046002",
    year = "2012"
}

@article{Aschieri:2013nda,
    author = "Aschieri, Paolo and Ferrara, Sergio",
    title = "{Constitutive relations and Schroedinger's formulation of nonlinear electromagnetic theories}",
    eprint = "1302.4737",
    archivePrefix = "arXiv",
    primaryClass = "hep-th",
    reportNumber = "CERN-PH-TH-2013-004",
    doi = "10.1007/JHEP05(2013)087",
    journal = "JHEP",
    volume = "05",
    pages = "087",
    year = "2013"
}

@article{He:2025ppz,
    author = "He, Song and Li, Yi and Ouyang, Hao and Sun, Yuan",
    title = "{$T\overline{T}$ Deformation: Introduction and Some Recent Advances}",
    eprint = "2503.09997",
    archivePrefix = "arXiv",
    primaryClass = "hep-th",
    month = "3",
    year = "2025"
}

@book{cour,
  author    = {R. Courant and D. Hilbert},
  title     = {Methods of Mathematical Physics, Vol. II},
  publisher = {Interscience},
  year      = {1962},
  note      = {See p. 93 and Chapters I and II \textit{passim}}
}

@article{Carrasco:2013qia,
    author = "Carrasco, John Joseph M. and Kallosh, Renata",
    title = "{Hidden Supersymmetry May Imply Duality Invariance}",
    eprint = "1303.5663",
    archivePrefix = "arXiv",
    primaryClass = "hep-th",
    reportNumber = "SU-ITP-13-03",
    month = "3",
    year = "2013"
}

@article{Russo:2024llm,
    author = "Russo, Jorge G. and Townsend, Paul K.",
    title = "{Causal self-dual electrodynamics}",
    eprint = "2401.06707",
    archivePrefix = "arXiv",
    primaryClass = "hep-th",
    doi = "10.1103/PhysRevD.109.105023",
    journal = "Phys. Rev. D",
    volume = "109",
    number = "10",
    pages = "105023",
    year = "2024"
}

@article{Babaei-Aghbolagh:2013hia,
    author = "Babaei-Aghbolagh, H. and Garousi, Mohammad R.",
    title = "{S-duality of tree-level S-matrix elements in D3-brane effective action}",
    eprint = "1304.2938",
    archivePrefix = "arXiv",
    primaryClass = "hep-th",
    doi = "10.1103/PhysRevD.88.026008",
    journal = "Phys. Rev. D",
    volume = "88",
    number = "2",
    pages = "026008",
    year = "2013"
}

@article{Garousi:2017fbe,
    author = "Garousi, Mohammad R.",
    title = "{Duality constraints on effective actions}",
    eprint = "1702.00191",
    archivePrefix = "arXiv",
    primaryClass = "hep-th",
    doi = "10.1016/j.physrep.2017.07.009",
    journal = "Phys. Rept.",
    volume = "702",
    pages = "1--30",
    year = "2017"
}

@article{Pavao:2022kog,
    author = "Pavao, Nicolas H.",
    title = "{Effective observables for electromagnetic duality from novel amplitude decomposition}",
    eprint = "2210.12800",
    archivePrefix = "arXiv",
    primaryClass = "hep-th",
    doi = "10.1103/PhysRevD.107.065020",
    journal = "Phys. Rev. D",
    volume = "107",
    number = "6",
    pages = "065020",
    year = "2023"
}

@article{BabaeiVelni:2019ptj,
    author = "Babaei Velni, Komeil and Babaei-Aghbolagh, H.",
    title = "{$S$-dual amplitude and $D_3$-brane couplings}",
    eprint = "1901.00198",
    archivePrefix = "arXiv",
    primaryClass = "hep-th",
    doi = "10.1103/PhysRevD.99.066007",
    journal = "Phys. Rev. D",
    volume = "99",
    number = "6",
    pages = "066007",
    year = "2019"
}

@article{Green:1996qg,
    author = "Green, Michael B. and Gutperle, Michael",
    title = "{Comments on three-branes}",
    eprint = "hep-th/9602077",
    archivePrefix = "arXiv",
    reportNumber = "DAMTP-96-12",
    doi = "10.1016/0370-2693(96)00331-0",
    journal = "Phys. Lett. B",
    volume = "377",
    pages = "28--35",
    year = "1996"
}

@article{Aschieri:2008ns,
    author = "Aschieri, Paolo and Ferrara, Sergio and Zumino, Bruno",
    title = "{Duality Rotations in Nonlinear Electrodynamics and in Extended Supergravity}",
    eprint = "0807.4039",
    archivePrefix = "arXiv",
    primaryClass = "hep-th",
    reportNumber = "CERN-PH-TH-2008-144, DESTA-UPO-08, LBNL-69756",
    doi = "10.1393/ncr/i2008-10038-8",
    journal = "Riv. Nuovo Cim.",
    volume = "31",
    number = "11",
    pages = "625--707",
    year = "2008"
}

@article{Babaei-Aghbolagh:2022MoxMax,
  author       = {H. Babaei-Aghbolagh and K. B. Velni and D. M. Yekta and H. Mohammadzadeh},
  title        = {Emergence of non-linear electrodynamic theories from $T\overline{T}$-like deformations},
  journal      = {Physics Letters B},
  volume       = {829},
  year         = {2022},
  pages        = {137079},
  doi          = {10.1016/j.physletb.2022.137079},
  archivePrefix = {arXiv},
  eprint       = {2202.11156},
  primaryClass = {hep-th}
}

@article{Bandos:2020hgy,
    author = "Bandos, Igor and Lechner, Kurt and Sorokin, Dmitri and Townsend, Paul K.",
    title = "{On p-form gauge theories and their conformal limits}",
    eprint = "2012.09286",
    archivePrefix = "arXiv",
    primaryClass = "hep-th",
    doi = "10.1007/JHEP03(2021)022",
    journal = "JHEP",
    volume = "03",
    pages = "022",
    year = "2021"
}

@article{Chen:2024vho,
    author = "Chen, Bin and Hou, Jue and Sun, Haowei",
    title = "{On self-dual Carrollian conformal nonlinear electrodynamics}",
    eprint = "2405.04105",
    archivePrefix = "arXiv",
    primaryClass = "hep-th",
    doi = "10.1007/JHEP08(2024)160",
    journal = "JHEP",
    volume = "08",
    pages = "160",
    year = "2024"
}

@article{Mkrtchyan:2019opf,
    author = "Mkrtchyan, Karapet",
    title = "{On Covariant Actions for Chiral $p-$Forms}",
    eprint = "1908.01789",
    archivePrefix = "arXiv",
    primaryClass = "hep-th",
    doi = "10.1007/JHEP12(2019)076",
    journal = "JHEP",
    volume = "12",
    pages = "076",
    year = "2019"
}

@article{Sorokin:2021tge,
    author = "Sorokin, Dmitri P.",
    title = "{Introductory Notes on Non-linear Electrodynamics and its Applications}",
    eprint = "2112.12118",
    archivePrefix = "arXiv",
    primaryClass = "hep-th",
    doi = "10.1002/prop.202200092",
    journal = "Fortsch. Phys.",
    volume = "70",
    number = "7-8",
    pages = "2200092",
    year = "2022"
}

@article{Aggarwal:2024yxy,
    author = "Aggarwal, Ankit and Ecker, Florian and Grumiller, Daniel and Vassilevich, Dmitri",
    title = "{Carroll-Hawking effect}",
    eprint = "2403.00073",
    archivePrefix = "arXiv",
    primaryClass = "hep-th",
    reportNumber = "TUW-24-02",
    doi = "10.1103/PhysRevD.110.L041506",
    journal = "Phys. Rev. D",
    volume = "110",
    number = "4",
    pages = "L041506",
    year = "2024"
}

@article{Stieberger:2024shv,
    author = "Stieberger, Stephan and Taylor, Tomasz R. and Zhu, Bin",
    title = "{Carrollian Amplitudes from Strings}",
    eprint = "2402.14062",
    archivePrefix = "arXiv",
    primaryClass = "hep-th",
    doi = "10.1007/JHEP04(2024)127",
    journal = "JHEP",
    volume = "04",
    pages = "127",
    year = "2024"
}

@article{Bagchi:2024unl,
    author = "Bagchi, Arjun and Banerjee, Aritra and Mondal, Saikat and Mukherjee, Debangshu and Muraki, Hisayoshi",
    title = "{Beyond Wilson? Carroll from current deformations}",
    eprint = "2401.16482",
    archivePrefix = "arXiv",
    primaryClass = "hep-th",
    doi = "10.1007/JHEP06(2024)215",
    journal = "JHEP",
    volume = "06",
    pages = "215",
    year = "2024"
}

@article{Liu:2024nkc,
    author = "Liu, Wen-Bin and Long, Jiang",
    title = "{Holographic dictionary from bulk reduction}",
    eprint = "2401.11223",
    archivePrefix = "arXiv",
    primaryClass = "hep-th",
    doi = "10.1103/PhysRevD.109.L061901",
    journal = "Phys. Rev. D",
    volume = "109",
    number = "6",
    pages = "L061901",
    year = "2024"
}

@article{Tadros:2024qlo,
    author = "Tadros, Poula and Kol\'a\v{r}, Ivan",
    title = "{Uniqueness of Galilean and Carrollian limits of gravitational theories and application to higher derivative gravity}",
    eprint = "2401.00967",
    archivePrefix = "arXiv",
    primaryClass = "gr-qc",
    doi = "10.1103/PhysRevD.109.084019",
    journal = "Phys. Rev. D",
    volume = "109",
    number = "8",
    pages = "084019",
    year = "2024"
}

@article{Russo:2024ptw,
    author = "Russo, Jorge G. and Townsend, Paul K.",
    title = "{Dualities of self-dual nonlinear electrodynamics}",
    eprint = "2407.02577",
    archivePrefix = "arXiv",
    primaryClass = "hep-th",
    doi = "10.1007/JHEP09(2024)107",
    journal = "JHEP",
    volume = "09",
    pages = "107",
    year = "2024"
}

@article{KOSYAKOV2020135840,
title = {Nonlinear electrodynamics with the maximum allowable symmetries},
journal = {Physics Letters B},
volume = {810},
pages = {135840},
year = {2020},
issn = {0370-2693},
doi = {https://doi.org/10.1016/j.physletb.2020.135840},
url = {https://www.sciencedirect.com/science/article/pii/S0370269320306432},
author = {B.P. Kosyakov}
}

@article{Hou:2022csf,
    author = "Hou, Jue",
    title = "{$ T\overline{T} $ flow as characteristic flows}",
    eprint = "2208.05391",
    archivePrefix = "arXiv",
    primaryClass = "hep-th",
    doi = "10.1007/JHEP03(2023)243",
    journal = "JHEP",
    volume = "03",
    pages = "243",
    year = "2023"
}

@article{Conti:2022egv,
    author = "Conti, Riccardo and Romano, Jacopo and Tateo, Roberto",
    title = "{Metric approach to a $ \mathrm{T}\overline{\mathrm{T}} $-like deformation in arbitrary dimensions}",
    eprint = "2206.03415",
    archivePrefix = "arXiv",
    primaryClass = "hep-th",
    doi = "10.1007/JHEP09(2022)085",
    journal = "JHEP",
    volume = "09",
    pages = "085",
    year = "2022"
}

@article{Soleng:1995kn,
    author = "Soleng, Harald H.",
    title = "{Charged black points in general relativity coupled to the logarithmic U(1) gauge theory}",
    eprint = "hep-th/9509033",
    archivePrefix = "arXiv",
    reportNumber = "CERN-TH-95-110, CERN-TH-95-110-REV",
    doi = "10.1103/PhysRevD.52.6178",
    journal = "Phys. Rev. D",
    volume = "52",
    pages = "6178--6181",
    year = "1995"
}

@article{BabaeiVelni:2016qea,
    author = "Babaei Velni, Komeil and Babaei-Aghbolagh, H.",
    title = "{On SL (2,R) symmetry in nonlinear electrodynamics theories}",
    eprint = "1610.07790",
    archivePrefix = "arXiv",
    primaryClass = "hep-th",
    doi = "10.1016/j.nuclphysb.2016.10.020",
    journal = "Nucl. Phys. B",
    volume = "913",
    pages = "987--1000",
    year = "2016"
}

@article{Mkrtchyan:2022ulc,
    author = "Mkrtchyan, Karapet and Svazas, Mantas",
    title = "{Solutions in Nonlinear Electrodynamics and their double copy regular black holes}",
    eprint = "2205.14187",
    archivePrefix = "arXiv",
    primaryClass = "hep-th",
    reportNumber = "Imperial-TP-KM-2022-2",
    doi = "10.1007/JHEP09(2022)012",
    journal = "JHEP",
    volume = "09",
    pages = "012",
    year = "2022"
}

@article{Ferko:2023wyi,
    author = "Ferko, Christian and Kuzenko, Sergei M. and Smith, Liam and Tartaglino-Mazzucchelli, Gabriele",
    title = "{Duality-invariant nonlinear electrodynamics and stress tensor flows}",
    eprint = "2309.04253",
    archivePrefix = "arXiv",
    primaryClass = "hep-th",
    doi = "10.1103/PhysRevD.108.106021",
    journal = "Phys. Rev. D",
    volume = "108",
    number = "10",
    pages = "106021",
    year = "2023"
}

@article{Gibbons:1995ap,
    author = "Gibbons, G W and Rasheed, D A",
    title = "{Sl(2,R) invariance of nonlinear electrodynamics coupled to an axion and a dilaton}",
    eprint = "hep-th/9509141",
    archivePrefix = "arXiv",
    reportNumber = "DAMTP-R-95-48",
    doi = "10.1016/0370-2693(95)01272-9",
    journal = "Phys. Lett. B",
    volume = "365",
    pages = "46--50",
    year = "1996"
}

@article{Levy-Leblond:1965,
    author = "Levy-Leblond, J.",
    title = "{Une nouvelle limite non-relativiste du groupe de poincaré}",
    //doi = "10.1063/1.1664490",
    journal = "Annales de l’IHP Physique théorique",
    volume = "3",
    pages = "1-12",
    year = "1965"
}

@article{SenGupta:1966qer,
    author = "Sen Gupta, N. D.",
    title = "{On an analogue of the Galilei group}",
    doi = "10.1007/BF02740871",
    journal = "Nuovo Cim. A",
    volume = "44",
    number = "2",
    pages = "512--517",
    year = "1966"
}

@article{Duval:2014uoa,
    author = "Duval, C. and Gibbons, G. W. and Horvathy, P. A. and Zhang, P. M.",
    title = "{Carroll versus Newton and Galilei: two dual non-Einsteinian concepts of time}",
    eprint = "1402.0657",
    archivePrefix = "arXiv",
    primaryClass = "gr-qc",
    doi = "10.1088/0264-9381/31/8/085016",
    journal = "Class. Quant. Grav.",
    volume = "31",
    pages = "085016",
    year = "2014"
}

@article{deBoer:2021jej,
    author = "de Boer, Jan and Hartong, Jelle and Obers, Niels A. and Sybesma, Watse and Vandoren, Stefan",
    title = "{Carroll Symmetry, Dark Energy and Inflation}",
    eprint = "2110.02319",
    archivePrefix = "arXiv",
    primaryClass = "hep-th",
    reportNumber = "NORDITA 2021-086",
    doi = "10.3389/fphy.2022.810405",
    journal = "Front. in Phys.",
    volume = "10",
    pages = "810405",
    year = "2022"
}

@article{Bergshoeff:2022qkx,
    author = "Bergshoeff, Eric A. and Gomis, Joaquim and Kleinschmidt, Axel",
    title = "{Non-Lorentzian theories with and without constraints}",
    eprint = "2210.14848",
    archivePrefix = "arXiv",
    primaryClass = "hep-th",
    doi = "10.1007/JHEP01(2023)167",
    journal = "JHEP",
    volume = "01",
    pages = "167",
    year = "2023"
}

@article{Chen:2021xkw,
    author = "Chen, Bin and Liu, Reiko and Zheng, Yu-fan",
    title = "{On Higher-dimensional Carrollian and Galilean Conformal Field Theories}",
    eprint = "2112.10514",
    archivePrefix = "arXiv",
    primaryClass = "hep-th",
    doi = "10.21468/SciPostPhys.14.5.088",
    journal = "SciPost Phys.",
    volume = "14",
    pages = "088",
    year = "2023"
}

@article{Banerjee:2020qjj,
    author = "Banerjee, Kinjal and Basu, Rudranil and Mehra, Aditya and Mohan, Akhila and Sharma, Aditya",
    title = "{Interacting Conformal Carrollian Theories: Cues from Electrodynamics}",
    eprint = "2008.02829",
    archivePrefix = "arXiv",
    primaryClass = "hep-th",
    doi = "10.1103/PhysRevD.103.105001",
    journal = "Phys. Rev. D",
    volume = "103",
    number = "10",
    pages = "105001",
    year = "2021"
}

@article{Bagchi:2016bcd,
    author = "Bagchi, Arjun and Basu, Rudranil and Kakkar, Ashish and Mehra, Aditya",
    title = "{Flat Holography: Aspects of the dual field theory}",
    eprint = "1609.06203",
    archivePrefix = "arXiv",
    primaryClass = "hep-th",
    doi = "10.1007/JHEP12(2016)147",
    journal = "JHEP",
    volume = "12",
    pages = "147",
    year = "2016"
}

@article{Chen:2023pqf,
    author = "Chen, Bin and Liu, Reiko and Sun, Haowei and Zheng, Yu-fan",
    title = "{Constructing Carrollian field theories from null reduction}",
    eprint = "2301.06011",
    archivePrefix = "arXiv",
    primaryClass = "hep-th",
    doi = "10.1007/JHEP11(2023)170",
    journal = "JHEP",
    volume = "11",
    pages = "170",
    year = "2023"
}

@article{Bandos:2020jsw,
    author = "Bandos, Igor and Lechner, Kurt and Sorokin, Dmitri and Townsend, Paul K.",
    title = "{A non-linear duality-invariant conformal extension of Maxwell's equations}",
    eprint = "2007.09092",
    archivePrefix = "arXiv",
    primaryClass = "hep-th",
    doi = "10.1103/PhysRevD.102.121703",
    journal = "Phys. Rev. D",
    volume = "102",
    pages = "121703",
    year = "2020"
}

@article{Ferko:2022cix,
    author = "Ferko, Christian and Sfondrini, Alessandro and Smith, Liam and Tartaglino-Mazzucchelli, Gabriele",
    title = "{Root-$T \bar T$ Deformations in Two-Dimensional Quantum Field Theories}",
    eprint = "2206.10515",
    archivePrefix = "arXiv",
    primaryClass = "hep-th",
    doi = "10.1103/PhysRevLett.129.201604",
    journal = "Phys. Rev. Lett.",
    volume = "129",
    number = "20",
    pages = "201604",
    year = "2022"
}

@article{Ferko:2022iru,
    author = "Ferko, Christian and Smith, Liam and Tartaglino-Mazzucchelli, Gabriele",
    title = "{On Current-Squared Flows and ModMax Theories}",
    eprint = "2203.01085",
    archivePrefix = "arXiv",
    primaryClass = "hep-th",
    doi = "10.21468/SciPostPhys.13.2.012",
    journal = "SciPost Phys.",
    volume = "13",
    number = "2",
    pages = "012",
    year = "2022"
}

@article{Henneaux:2021yzg,
    author = "Henneaux, Marc and Salgado-Rebolledo, Patricio",
    title = "{Carroll contractions of Lorentz-invariant theories}",
    eprint = "2109.06708",
    archivePrefix = "arXiv",
    primaryClass = "hep-th",
    doi = "10.1007/JHEP11(2021)180",
    journal = "JHEP",
    volume = "11",
    pages = "180",
    year = "2021"
}

@article{deBoer:2023fnj,
    author = "de Boer, Jan and Hartong, Jelle and Obers, Niels A. and Sybesma, Watse and Vandoren, Stefan",
    title = "{Carroll stories}",
    eprint = "2307.06827",
    archivePrefix = "arXiv",
    primaryClass = "hep-th",
    reportNumber = "NORDITA-2023-036",
    doi = "10.1007/JHEP09(2023)148",
    journal = "JHEP",
    volume = "09",
    pages = "148",
    year = "2023"
}

@article{Basu:2018dub,
    author = "Basu, Rudranil and Chowdhury, Udit Narayan",
    title = "{Dynamical structure of Carrollian Electrodynamics}",
    eprint = "1802.09366",
    archivePrefix = "arXiv",
    primaryClass = "hep-th",
    doi = "10.1007/JHEP04(2018)111",
    journal = "JHEP",
    volume = "04",
    pages = "111",
    year = "2018"
}

@article{He:2024yzx,
    author = "He, Song and Mao, Xin-Cheng",
    title = "{Irrelevant and marginal deformed BMS field theories}",
    eprint = "2401.09991",
    archivePrefix = "arXiv",
    primaryClass = "hep-th",
    doi = "10.1007/JHEP04(2024)138",
    journal = "JHEP",
    volume = "04",
    pages = "138",
    year = "2024"
}

@inproceedings{Gaillard:1997rt,
    author = "Gaillard, Mary K. and Zumino, Bruno",
    title = "{Nonlinear electromagnetic selfduality and Legendre transformations}",
    booktitle = "{A Newton Institute Euroconference on Duality and Supersymmetric Theories}",
    eprint = "hep-th/9712103",
    archivePrefix = "arXiv",
    reportNumber = "LBL-41110, LBNL-41110, UCB-PTH-97-58",
    pages = "33--48",
    month = "12",
    year = "1997"
}

@article{Chen:2023naw,
    author = "Chen, Bin and Hu, Zezhou",
    title = "{Bulk reconstruction in flat holography}",
    eprint = "2312.13574",
    archivePrefix = "arXiv",
    primaryClass = "hep-th",
    doi = "10.1007/JHEP03(2024)064",
    journal = "JHEP",
    volume = "03",
    pages = "064",
    year = "2024"
}

@article{Mehra:2024zqv,
    author = "Mehra, Aditya and Rathi, Hemant and Roychowdhury, Dibakar",
    title = "{Carrollian expansion of Born-Infeld electrodynamics}",
    eprint = "2401.06958",
    archivePrefix = "arXiv",
    primaryClass = "hep-th",
    doi = "10.1016/j.physletb.2024.139168",
    journal = "Phys. Lett. B",
    volume = "860",
    pages = "139168",
    year = "2025"
}

@article{Chen:2023esw,
    author = "Chen, Bin and Hu, Zezhou and Yu, Zhe-fei and Zheng, Yu-fan",
    title = "{Path-integral quantization of tensionless (super) string}",
    eprint = "2302.05975",
    archivePrefix = "arXiv",
    primaryClass = "hep-th",
    doi = "10.1007/JHEP08(2023)133",
    journal = "JHEP",
    volume = "08",
    pages = "133",
    year = "2023"
}

@article{Marsot:2023qlc,
    author = {Marsot, Lo\"\i{}c},
    title = "{Induced motions on Carroll geometries}",
    eprint = "2312.09924",
    archivePrefix = "arXiv",
    primaryClass = "gr-qc",
    doi = "10.1088/1361-6382/ad5cbc",
    journal = "Class. Quant. Grav.",
    volume = "41",
    number = "15",
    pages = "155010",
    year = "2024"
}

@article{Ecker:2023uwm,
    author = "Ecker, Florian and Grumiller, Daniel and Hartong, Jelle and P\'erez, Alfredo and Prohazka, Stefan and Troncoso, Ricardo",
    title = "{Carroll black holes}",
    eprint = "2308.10947",
    archivePrefix = "arXiv",
    primaryClass = "hep-th",
    reportNumber = "TUW-23-03",
    doi = "10.21468/SciPostPhys.15.6.245",
    journal = "SciPost Phys.",
    volume = "15",
    number = "6",
    pages = "245",
    year = "2023"
}

@article{Bergshoeff:2023vfd,
    author = "Bergshoeff, Eric A. and Campoleoni, Andrea and Fontanella, Andrea and Mele, Lea and Rosseel, Jan",
    title = "{Carroll fermions}",
    eprint = "2312.00745",
    archivePrefix = "arXiv",
    primaryClass = "hep-th",
    doi = "10.21468/SciPostPhys.16.6.153",
    journal = "SciPost Phys.",
    volume = "16",
    number = "6",
    pages = "153",
    year = "2024"
}

@article{Kasikci:2023zdn,
    author = "Kasikci, Oguzhan and Ozkan, Mehmet and Pang, Yi and Zorba, Utku",
    title = "{Carrollian supersymmetry and SYK-like models}",
    eprint = "2311.00039",
    archivePrefix = "arXiv",
    primaryClass = "hep-th",
    doi = "10.1103/PhysRevD.110.L021702",
    journal = "Phys. Rev. D",
    volume = "110",
    number = "2",
    pages = "L021702",
    year = "2024"
}

@article{Ciambelli:2023xqk,
    author = "Ciambelli, Luca",
    title = "{Dynamics of Carrollian scalar fields}",
    eprint = "2311.04113",
    archivePrefix = "arXiv",
    primaryClass = "hep-th",
    doi = "10.1088/1361-6382/ad5bb5",
    journal = "Class. Quant. Grav.",
    volume = "41",
    number = "16",
    pages = "165011",
    year = "2024"
}

@article{Mason:2023mti,
    author = "Mason, Lionel and Ruzziconi, Romain and Yelleshpur Srikant, Akshay",
    title = "{Carrollian amplitudes and celestial symmetries}",
    eprint = "2312.10138",
    archivePrefix = "arXiv",
    primaryClass = "hep-th",
    doi = "10.1007/JHEP05(2024)012",
    journal = "JHEP",
    volume = "05",
    pages = "012",
    year = "2024"
}

@article{Hao:2021urq,
    author = "Hao, Peng-xiang and Song, Wei and Xie, Xianjin and Zhong, Yuan",
    title = "{BMS-invariant free scalar model}",
    eprint = "2111.04701",
    archivePrefix = "arXiv",
    primaryClass = "hep-th",
    doi = "10.1103/PhysRevD.105.125005",
    journal = "Phys. Rev. D",
    volume = "105",
    number = "12",
    pages = "125005",
    year = "2022"
}

@article{Yu:2022bcp,
    author = "Yu, Zhe-fei and Chen, Bin",
    title = "{Free field realization of the BMS Ising model}",
    eprint = "2211.06926",
    archivePrefix = "arXiv",
    primaryClass = "hep-th",
    doi = "10.1007/JHEP08(2023)116",
    journal = "JHEP",
    volume = "08",
    pages = "116",
    year = "2023"
}

@article{Hao:2022xhq,
    author = "Hao, Peng-Xiang and Song, Wei and Xiao, Zehua and Xie, Xianjin",
    title = "{BMS-invariant free fermion models}",
    eprint = "2211.06927",
    archivePrefix = "arXiv",
    primaryClass = "hep-th",
    doi = "10.1103/PhysRevD.109.025002",
    journal = "Phys. Rev. D",
    volume = "109",
    number = "2",
    pages = "025002",
    year = "2024"
}

@article{Nguyen:2023miw,
    author = "Nguyen, Kevin",
    title = "{Carrollian conformal correlators and massless scattering amplitudes}",
    eprint = "2311.09869",
    archivePrefix = "arXiv",
    primaryClass = "hep-th",
    doi = "10.1007/JHEP01(2024)076",
    journal = "JHEP",
    volume = "01",
    pages = "076",
    year = "2024"
}

@article{Banerjee:2024jub,
    author = "Banerjee, Rabin and Bhattacharya, Soumya and Majhi, Bibhas Ranjan",
    title = "{Sengupta transformations and Carrollian relativistic theory}",
    eprint = "2403.02653",
    archivePrefix = "arXiv",
    primaryClass = "hep-th",
    doi = "10.1140/epjc/s10052-024-12959-4",
    journal = "Eur. Phys. J. C",
    volume = "84",
    number = "6",
    pages = "602",
    year = "2024"
}

@article{rainich1925electrodynamics,
  title={Electrodynamics in the general relativity theory},
  author={Rainich, George Yuri},
  journal={Transactions of the American Mathematical Society},
  volume={27},
  number={1},
  pages={106--136},
  year={1925},
  publisher={JSTOR}
}

@article{Gibbons:1995cv,
    author = "Gibbons, G. W. and Rasheed, D. A.",
    title = "{Electric - magnetic duality rotations in nonlinear electrodynamics}",
    eprint = "hep-th/9506035",
    archivePrefix = "arXiv",
    doi = "10.1016/0550-3213(95)00409-L",
    journal = "Nucl. Phys. B",
    volume = "454",
    pages = "185--206",
    year = "1995"
}

@article{Born:1933lls,
    author = "Born, Max and Infeld, L.",
    title = "{Electromagnetic mass}",
    doi = "10.1038/132970a0",
    journal = "Nature",
    volume = "132",
    number = "3347",
    pages = "970.1",
    year = "1933"
}

@article{Born:1934gh,
    author = "Born, M. and Infeld, L.",
    title = "{Foundations of the new field theory}",
    doi = "10.1098/rspa.1934.0059",
    journal = "Proc. Roy. Soc. Lond. A",
    volume = "144",
    number = "852",
    pages = "425--451",
    year = "1934"
}

@article{Babaei-Aghbolagh:2024uqp,
    author = "Babaei-Aghbolagh, H. and He, Song and Ouyang, Hao",
    title = "{Generalized $ T\overline{T} $-like deformations in duality-invariant nonlinear electrodynamic theories}",
    eprint = "2407.03698",
    archivePrefix = "arXiv",
    primaryClass = "hep-th",
    doi = "10.1007/JHEP09(2024)137",
    journal = "JHEP",
    volume = "09",
    pages = "137",
    year = "2024"
}

@inproceedings{Ferko:2024yhc,
    author = "Ferko, Christian and Luke Martin, Cian",
    title = "{Field-Dependent Metrics and Higher-Form Symmetries in Duality-Invariant Theories of Non-Linear Electrodynamics}",
    eprint = "2406.17194",
    archivePrefix = "arXiv",
    primaryClass = "hep-th",
    month = "6",
    year = "2024"
}

@article{Duval:2014uva,
    author = "Duval, C. and Gibbons, G. W. and Horvathy, P. A.",
    title = "{Conformal Carroll groups and BMS symmetry}",
    eprint = "1402.5894",
    archivePrefix = "arXiv",
    primaryClass = "gr-qc",
    doi = "10.1088/0264-9381/31/9/092001",
    journal = "Class. Quant. Grav.",
    volume = "31",
    pages = "092001",
    year = "2014"
}

@article{Duval:2014lpa,
    author = "Duval, C. and Gibbons, G. W. and Horvathy, P. A.",
    title = "{Conformal Carroll groups}",
    eprint = "1403.4213",
    archivePrefix = "arXiv",
    primaryClass = "hep-th",
    doi = "10.1088/1751-8113/47/33/335204",
    journal = "J. Phys. A",
    volume = "47",
    number = "33",
    pages = "335204",
    year = "2014"
}

@article{Hansen:2020hrs,
    author = "Hansen, Dennis and Jiang, Yunfeng and Xu, Jiuci",
    title = "{Geometrizing non-relativistic bilinear deformations}",
    eprint = "2012.12290",
    archivePrefix = "arXiv",
    primaryClass = "hep-th",
    reportNumber = "CERN-TH-2020-221",
    doi = "10.1007/JHEP04(2021)186",
    journal = "JHEP",
    volume = "04",
    pages = "186",
    year = "2021"
}

@article{Banerjee:2003bp,
    author = "Banerjee, Rabin",
    title = "{A Note on duality symmetry in nonlinear gauge theories}",
    eprint = "hep-th/0308162",
    archivePrefix = "arXiv",
    doi = "10.1016/j.physletb.2003.09.094",
    journal = "Phys. Lett. B",
    volume = "576",
    pages = "237--242",
    year = "2003"
}

@article{Hatsuda:1999ys,
    author = "Hatsuda, Machiko and Kamimura, Kiyoshi and Sekiya, Sayaka",
    title = "{Electric magnetic duality invariant Lagrangians}",
    eprint = "hep-th/9906103",
    archivePrefix = "arXiv",
    reportNumber = "KEK-TH-631, TOHO-FB-9963",
    doi = "10.1016/S0550-3213(99)00509-X",
    journal = "Nucl. Phys. B",
    volume = "561",
    pages = "341--353",
    year = "1999"
}

@article{Avetisyan:2021heg,
    author = "Avetisyan, Zhirayr and Evnin, Oleg and Mkrtchyan, Karapet",
    title = "{Democratic Lagrangians for Nonlinear Electrodynamics}",
    eprint = "2108.01103",
    archivePrefix = "arXiv",
    primaryClass = "hep-th",
    reportNumber = "Imperial-TP-KM-2021-02",
    doi = "10.1103/PhysRevLett.127.271601",
    journal = "Phys. Rev. Lett.",
    volume = "127",
    number = "27",
    pages = "271601",
    year = "2021"
}

@article{Banerjee:2022ocj,
    author = "Banerjee, Aritra and Dutta, Sudipta and Mondal, Saikat",
    title = "{Carroll fermions in two dimensions}",
    eprint = "2211.11639",
    archivePrefix = "arXiv",
    primaryClass = "hep-th",
    doi = "10.1103/PhysRevD.107.125020",
    journal = "Phys. Rev. D",
    volume = "107",
    number = "12",
    pages = "125020",
    year = "2023"
}

@article{Bagchi:2022eui,
    author = "Bagchi, Arjun and Banerjee, Aritra and Basu, Rudranil and Islam, Minhajul and Mondal, Saikat",
    title = "{Magic fermions: Carroll and flat bands}",
    eprint = "2211.11640",
    archivePrefix = "arXiv",
    primaryClass = "hep-th",
    doi = "10.1007/JHEP03(2023)227",
    journal = "JHEP",
    volume = "03",
    pages = "227",
    year = "2023"
}

@article{Banerjee:2022sza,
    author = "Banerjee, Aritra and Mehra, Aditya",
    title = "{Maximally symmetric nonlinear extension of electrodynamics with Galilean conformal symmetries}",
    eprint = "2206.11696",
    archivePrefix = "arXiv",
    primaryClass = "hep-th",
    doi = "10.1103/PhysRevD.106.085005",
    journal = "Phys. Rev. D",
    volume = "106",
    number = "8",
    pages = "085005",
    year = "2022"
}

@article{Babaei-Aghbolagh:2020kjg,
    author = "Babaei-Aghbolagh, H. and Babaei Velni, Komeil and Yekta, Davood Mahdavian and Mohammadzadeh, H.",
    title = "{$ T\overline{T} $-like flows in non-linear electrodynamic theories and S-duality}",
    eprint = "2012.13636",
    archivePrefix = "arXiv",
    primaryClass = "hep-th",
    reportNumber = "IPM/P-2020/066",
    doi = "10.1007/JHEP04(2021)187",
    journal = "JHEP",
    volume = "04",
    pages = "187",
    year = "2021"
}

@article{Babaei-Aghbolagh:2022itg,
    author = "Babaei-Aghbolagh, H. and Babaei Velni, Komeil and Yekta, Davood Mahdavian and Mohammadzadeh, H.",
    title = "{Manifestly SL(2, R) Duality-Symmetric Forms in ModMax Theory}",
    eprint = "2210.13196",
    archivePrefix = "arXiv",
    primaryClass = "hep-th",
    doi = "10.1007/JHEP12(2022)147",
    journal = "JHEP",
    volume = "12",
    pages = "147",
    year = "2022"
}

@article{Barbosa:2025smt,
    author = "Barbosa, Sergio and Fichet, Sylvain and de Souza, Lucas",
    title = "{On The Black Hole Weak Gravity Conjecture and Extremality in the Strong-Field Regime}",
    eprint = "2503.20910",
    archivePrefix = "arXiv",
    primaryClass = "hep-th",
    month = "3",
    year = "2025"
}

@article{Babaei-Aghbolagh:2025uoz,
    author = "Babaei-Aghbolagh, H. and Chen, Bin and He, Song",
    title = "{Root-$T\bar{T}$ Flows Unify 4D Duality-Invariant Electrodynamics and 2D Integrable Sigma Models}",
    eprint = "2507.22808",
    archivePrefix = "arXiv",
    primaryClass = "hep-th",
    month = "7",
    year = "2025"
}

@article{Murcia:2025psi,
    author = "Murcia, {\'A}ngel J.",
    title = "{Novel duality-invariant theories of electrodynamics}",
    eprint = "2507.16502",
    archivePrefix = "arXiv",
    primaryClass = "hep-th",
    month = "7",
    year = "2025"
}

@article{Russo:2025fuc,
    author = "Russo, Jorge G. and Townsend, Paul K.",
    title = "{Simplified self-dual electrodynamics}",
    eprint = "2505.08869",
    archivePrefix = "arXiv",
    primaryClass = "hep-th",
    month = "5",
    year = "2025"
}

\end{document}